\documentclass{aa}  

\usepackage{graphicx}
\usepackage[varg]{txfonts}
\usepackage{natbib}
\usepackage{soul}
\usepackage{caption}
\usepackage{subcaption}
\usepackage{multirow}
\usepackage{makecell}
\usepackage{array}
\usepackage{CJKutf8}
\usepackage[dvipsnames]{xcolor}
\usepackage{hyperref}
\definecolor{citeblue}{RGB}{35,75,125}
\hypersetup{
    breaklinks=true,
    colorlinks=true,
    linkcolor=black,   
    citecolor=black,   
    urlcolor=black      
}

\usepackage{enumerate}
\usepackage[inline]{enumitem}

\usepackage{amssymb}

\newcommand{\HI}{\ion{H}{I}}

\providecommand{\sorthelp}[1]{}

\begin{document} 

   \title{Statistical modeling of foreground Galactic cirrus for diffuse halo light measurement}

   \author{Qing Liu\inst{1}\thanks{e-mail: qliu@strw.leidenuniv.nl}, 
           Henk Hoekstra\inst{1}, 
           Ama\"el Ellien\inst{2}
          }

   \institute{Leiden Observatory, Leiden University, P.O. Box 9513, 2300 RA Leiden, The Netherlands
   \and
   OCA, P.H.C Boulevard de l’Observatoire CS 34229, 06304 Nice Cedex 4, France}
              
\titlerunning{cirrus modeling for diffuse halo light}
\authorrunning{Liu et al.}

\date{Received XXX; accepted XXX}

\abstract {Galactic cirrus, the diffuse Milky Way starlight scattered by interstellar dust, has become a major source of systematic uncertainty in measurements of extragalactic diffuse emission at ultra-low surface brightness levels. Proper modeling and correction of this foreground is crucial to unlock the full potential of next-generation deep imaging surveys for studies of faint diffuse emission such as galactic stellar halos, ultra-diffuse galaxies, and intracluster light (ICL).
We develop a statistical approach to reconstruct the 2D distribution of the Galactic cirrus while preserving the underlying diffuse extragalactic emission from stellar halos.
We combine the broad dust morphology traced by far-infrared emission with multiscale structural information extracted from the optical images. The method synthesizes plausible cirrus structures through iteratively refining the foreground, the smooth diffuse light model, and the residual sky. 
The performance is assessed through injection--recovery experiments using \textit{Euclid} VIS observations of the Abell~2390 cluster, which show substantial improvements in recovered stellar halo photometry, although residual biases depend on the local foreground structure and sky estimation.
We apply the method to the central Abell~2390 field to correct for a prominent cirrus complex near its brightest cluster galaxy (BCG), revealing a smoother BCG+ICL distribution with a steeper outer light profile and reduced contamination in the integrated light.
Our results demonstrate the potential of statistical modeling of Galactic cirrus to improve diffuse light photometry in deep wide-field imaging surveys.
}

\bibpunct{(}{)}{;}{a}{}{,} 


\maketitle
%

\section{Introduction}

The faint outskirts of galactic stellar halos provide unique insights about their formation and evolution. For massive galaxies, extended stellar halos serve as archaeological records of past mergers and accretion of satellite galaxies \citep[e.g.,][]{Bullock2005,Pillepich2014,DSouza2018}. Their surface brightness profiles, colors, and stellar mass fractions constrain the contribution of stellar accretion and the mass assembly history \citep[e.g.,][]{Merritt2016,DSouza2018,Huang2018a,Gilhuly2022}. 
On the scale of galaxy clusters, the intracluster light (ICL), which consists of stars stripped from infalling satellite galaxies, traces the dynamical assembly of the cluster and the growth of its brightest cluster galaxy \citep[BCG; e.g., ][]{Mihos2005,Montes2018,Montes2022}.
Comparing these observations with cosmological simulations provides valuable tests of the hierarchical galaxy formation models \citep[e.g.,][]{RodriguezGomez2016,Pillepich2018,MontenegroTaborda2023}.

At lower stellar masses, dwarf satellite galaxies and ultra-diffuse galaxies (UDGs) provide a complementary view where the evolution is less dominated by mergers \citep{Simon2019}. 
UDGs are particularly intriguing because they have stellar masses typical of dwarf galaxies but can have sizes comparable to those of much more massive galaxies \citep[e.g.,][]{Koda2015,vanDokkum2015}. 
The structural properties of these galaxies encode information about past star formation, stellar feedback, and environmental effects, providing important clues to understanding their formation mechanisms and physical processes that shape these systems \citep[e.g.,][]{ElBadry2016,Venhola2017,KadoFong2020}.

Deep imaging has now reached surface brightness limits in the optical that are fainter than 30 mag/arcsec$^2$, opening new opportunities for investigating the low surface brightness universe \citep[e.g.,][]{Abraham2014,Merritt2016,Trujillo2016,Gilhuly2022,Zaritsky2024,Cuillandre2025a,laRosa2026}. Together with a wide field of view, this depth enables the discovery and characterization of diffuse galaxies and faint stellar structures around galaxies over a large sample size \citep[e.g.,][]{Zaritsky2019,Marleau2025,Heesters2025,Sola2025}. At these levels, however, accurate measurements of this diffuse component require exquisite control of systematic errors such as sky background subtraction, wide-angle point spread function (PSF), instrumental stray light, and imperfect flat-fielding \citep[e.g.,][]{Slater2009,Sandin2014,Duc2015,FliriTrujillo2016,InfanteSainz2020,Liu2022,Liu2023,Watkins2024,Cuillandre2025a}.

Among all the systematics, Galactic cirrus is one of the most challenging astrophysical components to low surface brightness measurements. At optical wavelengths, this emission arises predominantly from Galactic starlight scattered by dust in the diffuse interstellar medium, producing diffuse Galactic light \citep{Sandage1976}. Its filamentary and patchy morphology exhibits brightness fluctuations across a broad range of angular scales, overlapping with those of stellar halos, tidal tails, diffuse galaxies, and ICL \citep[e.g.,][]{Mihos2017,Kluge2025,Liu2025a,MartínezDelgado2025,Poliakov2026}.
Cirrus can therefore obscure faint galaxies, resemble tidal structures, bias the sky estimates, and add excess light to source photometry \citep{Cortese2010,Mihos2019,Duc2015,Bilek2020}. Unlike many other systematics, this foreground remains after careful low surface brightness optimized data reduction and processing. Increasing the depth does not beat down these effects, since fainter cirrus structures will emerge. As a result, accurate photometry of the faint outskirts of galaxies requires a reliable estimate of this foreground superposed on the diffuse stellar emission.

The use of multiband colors and infrared dust emission provides useful means for identifying Galactic cirrus. Its optical colors can help distinguish cirrus from old stellar populations \citep{Roman2020,Smirnov2023,Liu2025a}, while the correlation between optical and far-infrared (FIR) emission connects dust scattered light to thermal emission from the same dust \citep[e.g.,][]{Laureijs1987,Guhathakurta1989,Witt2008,Ienaka2013,Zhang2023,Zhao2024}. In particular, \cite{Liu2025a} have shown that combining the color information calibrated with thermal dust maps with its optical morphology can facilitate the detection of faint galaxies in cirrus-rich fields. 
For overlapping cirrus and extragalactic light, however, spectral information alone does not specify how the two components are spatially distributed. 

Although optical and FIR cirrus show similar large-scale morphology, it is difficult to translate FIR observations  directly into optical templates at comparable resolution. The optical-to-FIR ratio has intrinsic scatter due to variations in dust properties and illumination geometry (\citealt{Ienaka2013,Zhang2023,Zhao2024,Kaboud2026}; Liu et al. 2026 in prep), while the broader FIR beam blends resolved optical filaments and fluctuations into  smoother structures. Unresolved galaxies and cosmic infrared background (CIB) anisotropies introduce additional FIR fluctuations that are not associated with Galactic dust \citep{Viero2013}. 
FIR emission can therefore provide a valuable guide to the dust distribution, whereas the optical cirrus surrounding a target could offer more direct information about its fine-scale morphology.

The spatial coherence of cirrus structures is driven by multiple physical processes that shape the diffuse interstellar medium (ISM). Observations of {\HI} and polarized dust emission have revealed extended filamentary structures whose orientations are correlated with the magnetic field \citep{Clark2014,PlanckCollaborationXXXII2016}, while numerical simulations have shown how turbulence, thermal instability, compression, and magnetic fields can jointly shape the density structure of the diffuse atomic medium \citep[e.g.,][]{Saury2014,InoueInutsuka2016}. Optical cirrus reflects this structural coherence over a broad range of angular scales \citep{MivilleDeschenes2016,Liu2025b}. 

Cirrus morphology can be characterized through statistical measures of its structure, such as fractal geometry, power spectra, and higher-order statistics \citep{Marchuk2021,MivilleDeschenes2016,Liu2025b}. 
In particular, the wavelet scattering transform (WST) provides a compact description of non-Gaussian structures across different scales and orientations of the diffuse ISM \citep{Allys2019}. WST statistics have been used to characterize optical cirrus and to generate fields with similar morphological patterns \citep{Liu2025b}. 
More recently, wavelet-based techniques have also been applied directly to the separation of the diffuse component. In the analysis of the Abell 2390 ICL, \citet{Ellien2025} used the Detection Algorithm for Intracluster light Studies \citep[DAWIS;][]{Ellien2021} to model and remove the large-scale cirrus component in the \textit{Euclid} NISP images. Wavelet phase harmonics methods have been used to separate Galactic dust from CIB anisotropies in Herschel and Planck observations \citep{Auclair2024,Sinha2026}. 
These statistical descriptions provide a basis for reconstructing cirrus where it overlaps diffuse sources, thereby mitigating its bias on measurements of the underlying light.

In this paper, we investigate whether statistical modeling of the Galactic cirrus can improve photometry of diffuse stellar emission in optical imaging. 
We reconstruct the cirrus foreground across a diffuse target source by combining the observed cirrus in its surroundings with an external FIR dust tracer and constraints on its multiscale morphology based on WST statistics. 
We assess this method by injecting halos with known properties into \textit{Euclid} VIS images that include cirrus and measuring how accurately their radial profiles and integrated light are recovered. We then apply this method to the central Abell~2390 field, where cirrus strongly affects the BCG and ICL measurements in the image. Additional experiments with synthetic cirrus, for which both the foreground and injected halo are known, are presented in Appendix~\ref{app:mock_recovery}. 
Throughout this paper, we assume a standard flat
Lambda cold-dark-matter ($\Lambda$CDM) cosmology with $H_0=70$ km\,s$^{-1}$\,Mpc$^{-1}$ and $\Omega_m=0.3$.

\section{Dataset} \label{sec:data}

\subsection{\textit{Euclid} VIS imaging of Abell 2390}
\label{sec:data_euclid}

\begin{figure*}[!htbp]
\centering
  \resizebox{\hsize}{!}{\includegraphics{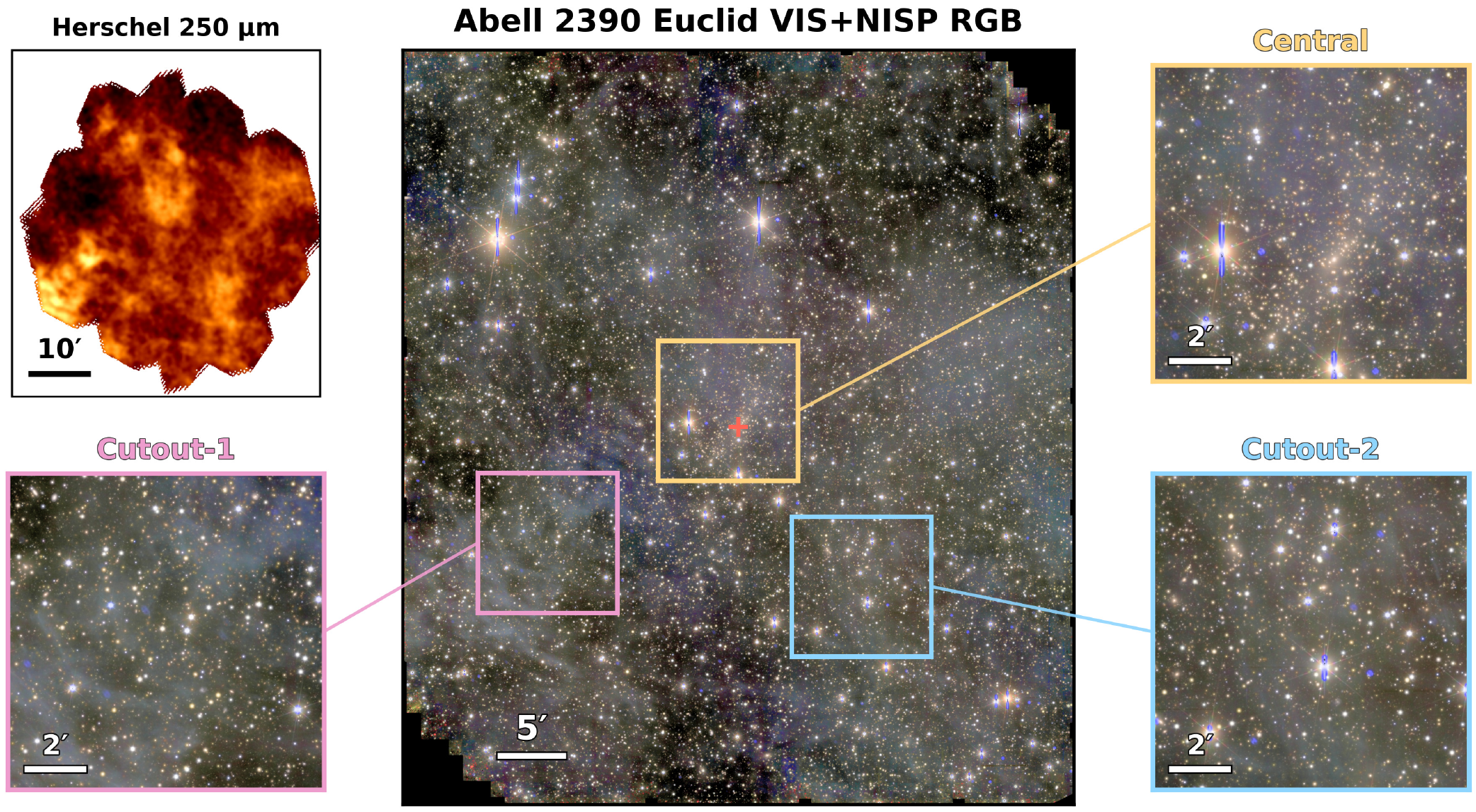}}
  \caption{\textbf{Overview of the \textit{Euclid} Abell 2390 field and FIR dust tracer.} The main canvas shows an RGB color composite of the \textit{Euclid} ERO Abell~2390 field (B: $I_{\rm E}$; G: $Y_{\rm E}$; R: $J_{\rm E}$). The red cross marks the BCG. Insets show the regions used for the halo injection experiments and the central region containing the BCG and ICL. The top left panel shows the cleaned, low-pass-filtered Herschel/SPIRE $250\,\mu{\rm m}$ image. Galactic cirrus extends across the \textit{Euclid} field and is particularly prominent in VIS; the correspondence with the FIR emission is apparent on broad scales, while finer structure is resolved in the optical image.}
\label{fig:A2390_rgb}
\end{figure*}

Abell~2390 is a rich massive cool-core galaxy cluster ($M_{200} \approx 2 \times 10^{15}M_\odot$) at redshift of 0.228 \citep{EROLensData,Schrabback25}.
It has been widely studied to understand galaxy evolution in dense environments \citep{Yee1996,Abraham1996,BaloghMorris2000}. The BCG is located at ${\rm RA}=21^{\rm h}53^{\rm m}36.83^{\rm s}$ and ${\rm Dec}=+17^{\circ}41'43.73''$ (J2000).
Abell~2390 was observed as part of the \textit{Euclid} ERO program \citep{EROcite} with three reference observing sequences, reaching three times the nominal \textit{Euclid} Wide Survey exposure time over a field of approximately $0.75\,{\rm deg}^{2}$ \citep{Cuillandre2025a,Ellien2025}. 
The observations include the broad optical $I_{\rm E}$ band from the VIS instrument and the $Y_{\rm E}$, $J_{\rm E}$, and $H_{\rm E}$ near-infrared (NIR) bands from NISP. Figure~\ref{fig:A2390_rgb} presents the full field as a color composite. The insets show regions used for the cirrus reconstruction experiments and the central BCG+ICL analysis.

We use the low surface brightness optimized VIS stack, which preserves large-scale emission that conventional local sky subtraction would remove \citep{Cuillandre2025a}. The native VIS images have a pixel scale of $0 \farcs 1$. For the Abell~2390 field, the ERO dataset reaches a $1\sigma$ low surface brightness limit of 30.42 mag/arcsec$^{2}$ on $10\arcsec \times 10 \arcsec$ scales \citep{Cuillandre2025a}. The final VIS stack adopts an AB photometric zero point of $30.13$. For the analysis in this work, we use the $3\times3$ binned VIS product with a pixel scale of $0 \farcs 3$ to detect and mask sources; we do a further $3\times3$ binning for the cirrus reconstruction. At the redshift of Abell~2390, $1\arcsec$ corresponds to a physical
scale of 3.65~kpc with the adopted cosmology.

Galactic cirrus is visible across the Abell 2390 field and is particularly prominent in the VIS image. \citet{Ellien2025} showed that this foreground complicates diffuse light measurements in VIS and therefore based their BCG+ICL analysis primarily on the less contaminated NISP images, after cirrus modeling with DAWIS. Here we treat the strong foreground in the VIS image as a test case for recovering diffuse light in the presence of strong cirrus contamination. The NISP images are not used to constrain the cirrus reconstruction presented in this work.

\subsection{Herschel $250\,\mu{\rm m}$ imaging}
\label{sec:data_herschel}

We use FIR imaging at $250\,\mu{\rm m}$ from the Spectral and Photometric Imaging Receiver \citep[SPIRE;][]{Griffin2010} on board the Herschel Space Observatory. The data were retrieved from the Herschel Science Archive and are calibrated in MJy/sr. The native SPIRE $250\,\mu{\rm m}$ band has a pixel scale of $6\arcsec$ and a beam FWHM of $18\arcsec$, and provides the highest angular resolution of the three SPIRE bands. Similar Herschel products have been used to trace Galactic cirrus in deep optical imaging \citep{Mihos2017,Zhang2023,Liu2025b,Kaboud2026}.

The FIR image contains emission not associated with Galactic cirrus, primarily from compact or extended infrared sources and the CIB anisotropies. We therefore preprocessed the map before using it as a dust prior. Following the approach of \cite{Liu2025b}, compact sources were detected in the Herschel image down to S/N 2.5 and masked. We then applied a low-pass filter to suppress residual small-scale fluctuations dominated by unresolved sources and CIB anisotropies. The resulting map has an effective angular resolution of approximately $36''$. We refer to this cleaned and smoothed $250\,\mu{\rm m}$ image as the FIR \emph{backbone} throughout this paper.

The FIR backbone is not treated as a pixel-by-pixel prediction of the optical cirrus. We use it only to trace scales where optical and FIR cirrus are closely related, though not strictly proportional. The fine resolution structure is constrained as described in Sec.~\ref{sec:method}. The top left inset in Fig.~\ref{fig:A2390_rgb} shows the FIR data, which illustrates the large-scale correspondence between optical and FIR cirrus and the large difference in angular resolution.

\section{Statistical reconstruction of Galactic cirrus} \label{sec:method}

\begin{figure*}[!htbp]
\centering
  \resizebox{\hsize}{!}{\includegraphics{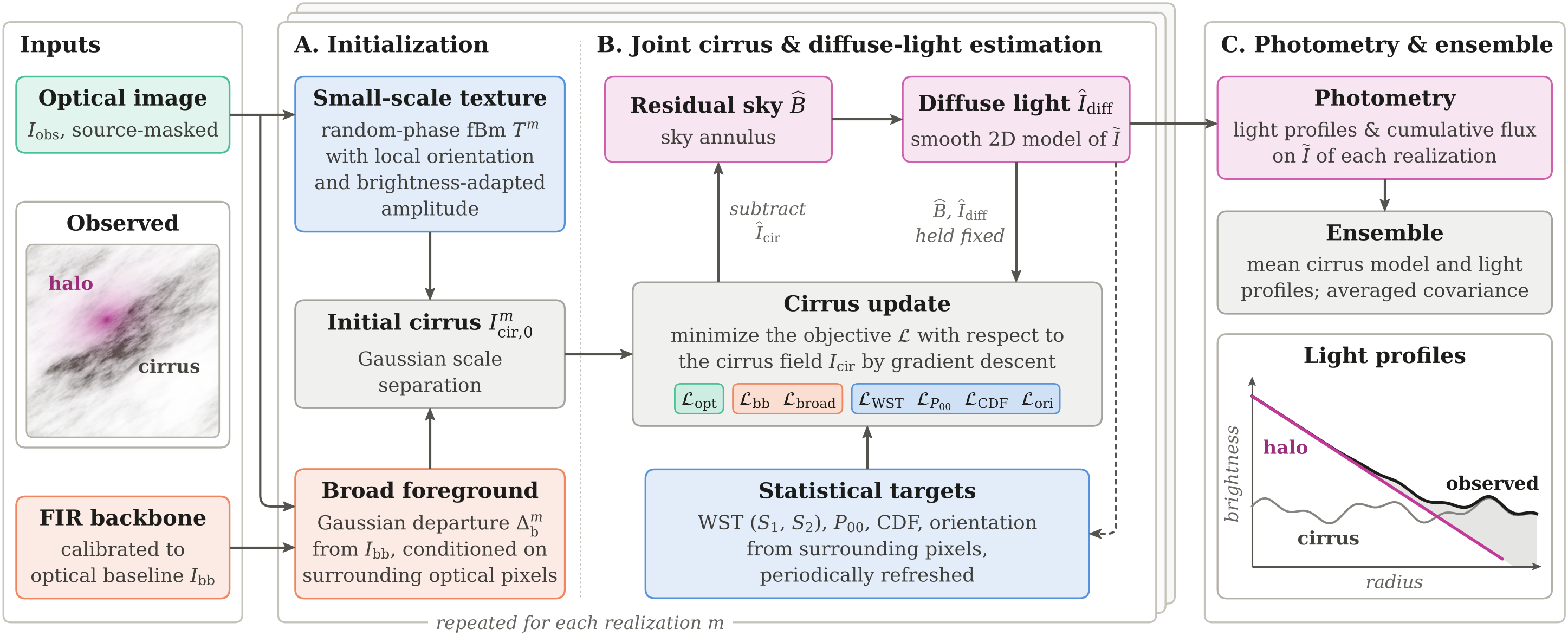}}
  \caption{\textbf{Cirrus and diffuse halo light modeling workflow.}
  The cirrus foreground is initialized from the FIR backbone and source-masked optical pixels (A). It is optimized jointly with the residual sky and the diffuse light (B). The photometry on the cirrus-subtracted images is averaged over the realizations (C). Steps A and B are repeated for each realization. The dashed arrow indicates the refresh of the statistical targets after updating the diffuse light model. The observed image and the profile plots are schematic.}
\label{fig:workflow}
\end{figure*}

The main goal is to infer the optical cirrus distribution extending across a diffuse halo. We defer more complicated cases (e.g., tidal features) to future work. We combine three sources of information: the broad dust distribution traced by the FIR backbone, the multiscale morphology of cirrus measured from the surrounding pixels, and the source-masked pixels that constrain the data likelihood. The final photometric measurement is performed on the cirrus subtracted image. 
Figure~\ref{fig:workflow} summarizes the workflow.

\subsection{Inference framework}
\label{sec:method_framework}

At the delivered pixel resolution, the observed intensity is described as
\begin{equation}
    I_{\rm obs}(\boldsymbol{x}) =
    I_{\rm cir}(\boldsymbol{x}) +
    I_{\rm diff}(\boldsymbol{x}) + 
    B + \epsilon(\boldsymbol{x}),
    \label{eq:forward_model}
\end{equation}
where \(I_{\rm cir}\) is the Galactic cirrus foreground, \(I_{\rm diff}\) is the diffuse stellar emission associated with the target, \(B\) is a residual sky level, and \(\epsilon\) represents instrumental noise. We define the cirrus-subtracted image as
\begin{equation}
    \widetilde{I}(\boldsymbol{x})
    \equiv
    I_{\rm obs}(\boldsymbol{x}) -
    \widehat{I}_{\rm cir}(\boldsymbol{x}) -
    \widehat{B}\,,
    \label{eq:cirrus_sub}
\end{equation}
where $\widehat{I}_{\rm cir}(\boldsymbol{x})$ and $\widehat{B}$ are the inferred cirrus model and sky level.
The resulting $\widetilde{I}$ contains the diffuse target emission together with observational noise and other residual patterns.

Compact and extended sources, stellar diffraction spikes, and image artifacts remain masked throughout the analysis. Around the target, we additionally define a central core region based on its effective radius ($R_{\rm e}$) that is withheld from the likelihood calculation for $I_{\rm cir}$. The pixels within this core remain available for estimating $I_{\rm diff}$. The fraction of contamination from cirrus at the bright core of the target is typically small.

Outside the core, cirrus and diffuse stellar light are constrained simultaneously. The cirrus model $I_{\rm cir}$ is required to remain consistent with the FIR backbone and with the multiscale statistics of the surrounding cirrus, whereas the diffuse halo $I_{\rm diff}$ is represented as a smooth 2D stellar light distribution. We alternate between these two estimates during the reconstruction, so that the cirrus model is constrained by the surrounding pixels while the diffuse light model reduces the imprint of the target on the foreground constraints.

\subsection{Conditional cirrus initialization}
\label{sec:method_init}

We initialize the optical cirrus distribution using information from two sources (step A in Fig.~\ref{fig:workflow}). The FIR data provide a broad baseline for the foreground, while the surrounding optical pixels constrain local
departures from it. For each target, we define an ellipse enclosing most of the expected diffuse light and exclude the pixels inside this ellipse from the initial conditioning. The size and geometry of the ellipse are specified for each experiment later. We denote the FIR-informed optical baseline as
$I_{\rm bb}(\boldsymbol{x})$, where ``bb'' refers to the backbone.

We describe the broad optical--FIR discrepancy as
\begin{equation}
\Delta_{\rm b}(\boldsymbol{x}) \equiv
I_{\rm cir}(\boldsymbol{x})-I_{\rm bb}(\boldsymbol{x}),
\label{eq:broad}
\end{equation}
and model it as a stationary Gaussian field with covariance $\mathbf{K}_{\rm b}$. By conditioning this field on the optical residuals outside the target ellipse, we estimate how the broad foreground structure continues into the excluded region.\footnote{This approach is conceptually similar to the Local Pixelwise Infilling (LPI) method of \citet{Saydjari2022}. Here we adopt a parametric covariance instead of evaluating it non-parametrically from neighboring image patches because of the aggressive source masking.}

The conditional mean of $\Delta_{\rm b}$ provides the mean expected broad foreground under this model, but does not capture the dispersion of structures. We therefore generate independent realizations conditioned on the same exterior pixels. These realizations remain closely constrained where observations are available, while allowing different structures within the target region. The covariance model for $\Delta_{\rm b}$ and conditional sampling are detailed in Appendix~\ref{app:init_broad}.

Small-scale cirrus structure is generated independently from random-phase Gaussian fields with power-law spectra, following the fractional Brownian motion (fBm) framework commonly used to describe interstellar emission \citep{MivilleDeschenes2003}.
Since an isotropic fBm does not reproduce the anisotropy and non-Gaussian morphology of real cirrus, we adapt its fluctuation power and preferred orientations using measurements from the surrounding optical data. We also modulate the fluctuation amplitude according to local cirrus brightness, motivated by the observed relation between FIR cirrus brightness and brightness fluctuation \citep{MivilleDeschenes2007}. Here, this dependence is measured directly from the data using unmasked pixels.
We denote the resulting stochastic texture for realization $m$ as $T^m$, with independent phases between realizations.

The conditioned broad field and stochastic texture are combined through a Gaussian scale separation:
\begin{equation}
\begin{aligned}
I_{{\rm cir},0}^{m}
={}&
\underbrace{
I_{\rm bb}+G_{\sigma_{\rm s}}\left[\Delta_{\rm b}^{m}\right]
}_{\text{Broad foreground}} \, +
\underbrace{
T^{m}-G_{\sigma_{\rm s}}\left[T^{m}\right]
}_{\text{Small-scale fluctuations}},
\end{aligned}
\label{eq:cirrus_initial}
\end{equation}
where $G_{\sigma_{\rm s}}$ denotes Gaussian smoothing with $\sigma_{\rm s}=12$ pixels. The first two terms determine the low-frequency foreground, while the last two terms describe the independently generated small-scale stochastic texture.

For each field, we construct several initial cirrus maps with independent broad-field realizations and texture phases, keeping the masks, FIR calibration, covariance model, and statistical constraints fixed. Note that the target ellipse is used only to exclude pixels from this initial conditioning.

\subsection{Multiscale morphology constraints}
\label{sec:method_stats}

We characterize the cirrus morphology using WST statistics measured from the optical image. For a family of wavelets $\psi_{j,\theta}$, the first- and second-order scattering coefficients are:
\begin{align}
S_1(j,\theta) &=
\left\langle
\left| I_{\rm cir}\star\psi_{j,\theta} \right|
\right\rangle ,
\label{eq:wst_s1}
\\
S_2(j_1,\theta_1,j_2,\theta_2) &=
\left\langle
\left| 
\left| I_{\rm cir}\star\psi_{j_1,\theta_1} \right| \star\psi_{j_2,\theta_2} 
\right| 
\right\rangle ,
\label{eq:wst_s2}
\end{align}
where $\langle\cdot\rangle$ denotes spatial averaging and we retain the physically informative indices with $j_2>j_1$. We compute the WST using the \texttt{scattering} package \citep{ChengMenard2020}. For a $10\arcmin \times 10 \arcmin$ cutout field, we adopt five spatial scales ($J=5$) and eight orientations ($\Theta=8$). In brief, $S_1$ describes the fluctuation amplitude across scale and orientation, while $S_2$ captures couplings between structures on different scales and directions. These higher-order statistics retain non-Gaussian morphological information. WST-based statistics have been used to characterize turbulent ISM structure and neutral-gas morphology \citep{Allys2019,LeiClark2023,Lei2025,Liu2025b}.

The statistical targets are measured from source-masked pixels surrounding the target after subtracting the current estimates of diffuse stellar light and residual sky, and are periodically refreshed as the separation updates. Instrumental noise and the mask response are accounted for using independent noise realizations and stochastic cirrus references. At the smallest wavelet scales, where the measured coefficients are more sensitive to noise and missing pixels, the WST targets are blended with the independent stochastic texture after accounting for instrumental noise and mask response (see Appendix~\ref{app:method_stats}).

We complement the WST with several statistics that describe other aspects of the foreground: the $P_{00}$ term measures directional wavelet energy across spatial scales, the cumulative distribution function (CDF) characterizes high-frequency fluctuations in different cirrus-brightness regimes, and local angular-energy statistics constrain the preferred structural orientation where the anisotropy is well measured. These statistics describe the amplitude, distribution, anisotropy, and cross-scale organization of the cirrus structures.

\subsection{Cirrus reconstruction objective}
\label{sec:method_objective}

Starting from each conditional initialization $I_{{\rm cir},0}^{m}$,
we reconstruct the cirrus field by combining the observed optical pixel intensities, the coarse foreground informed by FIR data, and the multiscale morphology
measured from the surrounding optical cirrus. The total objective is:
\begin{equation}
\begin{split}
\mathcal{L} ={}&
\lambda_{\rm opt}\mathcal{L}_{\rm opt}
+ \lambda_{\rm broad}\mathcal{L}_{\rm broad}
+ \lambda_{\rm bb}\mathcal{L}_{\rm bb}
\\
& + \lambda_{\rm WST}\mathcal{L}_{\rm WST}
+ \lambda_{P_{00}}\mathcal{L}_{P_{00}}
+ \lambda_{\rm CDF}\mathcal{L}_{\rm CDF}
+ \lambda_{\rm ori}\mathcal{L}_{\rm ori}.
\label{eq:objective}
\end{split}
\end{equation}
The $\lambda$ coefficients control the relative strengths of the normalized constraints and are fixed for each experiment.
The latter four statistical terms constrain the multiscale structure, fluctuation distribution, and local orientations of cirrus, as described in Sec.~\ref{sec:method_stats}. More details are given in Appendix~\ref{app:method_stats}.

The two FIR-informed terms serve different purposes.
The backbone term $\mathcal{L}_{\rm bb}$ compares the cirrus model with the calibrated FIR template after convolution with the effective FIR beam. Its covariance accounts for measurement uncertainty, calibration uncertainty, and scatter in the optical--FIR relation. 
The broad-field term $\mathcal{L}_{\rm broad}$ instead regularizes the large-scale optical--FIR discrepancy $\Delta_{\rm b}$, favoring departures from the backbone consistent with the covariance model. More details about the FIR-informed likelihoods are described in Appendix~\ref{app:method_FIR_lld}.

The optical term $\mathcal{L}_{\rm opt}$ constrains the cirrus intensity using the residual between the observed image and the cirrus, diffuse light, and sky models.
It is evaluated on source-masked pixels outside a central exclusion of the target, so that the remaining pixels in the initialization ellipse also contribute. A covariance model accounts for correlated instrumental noise and additional uncertainty in the overlapping cirrus and diffuse-light components. The detailed objective functions are described in Appendix~\ref{app:method_optical_lld}.

\subsection{Joint cirrus and diffuse light estimation}
\label{sec:method_halo}

The cirrus and diffuse halo light are estimated jointly, since an error in either component can bias the inference of the other. In practice, we alternate between cirrus optimization and diffuse-light fitting (step B in Fig.~\ref{fig:workflow}). After each block of cirrus updates, we subtract the current cirrus estimate $\widehat{I}_{\rm cir}$ from the observed image $I_{\rm obs}$, re-estimate the residual sky $\widehat{B}$ from an outer annulus and subtract it, and generate a cirrus-and-sky subtracted image. We then fit a smooth 2D diffuse-light model $\widehat{I}_{\rm diff}$ to the remaining emission $\widetilde{I}$.

The diffuse light model is constructed from elliptical isophotes fitted with \texttt{photutils} \citep{Bradley2016}.\footnote{We use elliptical isophotes to illustrate the modeling of smooth halo light, although this choice is not inherent to the reconstruction method. An isophote model cannot describe diffuse emission such as ICL substructures. One could in principle incorporate a more complicated model into the same iterative framework.}
The center is fixed, while ellipticity and orientation are allowed to vary where the isophotes provide sufficient observational support. The geometry is smoothed as a function of radius and held fixed beyond the last reliable isophote, and the amplitude is kept non-negative. When a reliable isophote sequence cannot be recovered, we instead adopt a fixed conditional geometry based on the inner light distribution and fit only its radial intensity profile.

The updated diffuse light model and residual sky are held fixed during the next cirrus optimization interval. The statistical cirrus targets are periodically re-evaluated from the image after subtracting the current diffuse light and sky estimates. After the final cirrus update, the diffuse light model and residual sky are fitted once more using the final cirrus model. Further implementation details are provided in
Appendix~\ref{app:method_optimization}.

\section{Recovering injected halos in \textit{Euclid} imaging} \label{sec:halo}

With the cirrus modeling approach introduced in Sec.~\ref{sec:method}, we investigate how reliably can we recover the light of a diffuse halo heavily contaminated by the cirrus foreground.
We address this through injection--recovery experiments that place known low surface brightness sources into \textit{Euclid} images. 
We first describe the experimental design and then examine two representative cases spanning an extended massive halo envelope and a UDG-like diffuse source.

\subsection{Injection--recovery experiment design}
\label{sec:halo_design}

We inject model diffuse sources into the observed \textit{Euclid} VIS images. The Galactic foreground, source masks, instrumental noise, and detector artifacts are inherited from the observations. 
Because the majority of Galactic cirrus is in the diffuse ISM regime, we neglect the weak dust reddening and extinction effects that the foreground would impose on the source.

We consider two different diffuse halo models. In Cutout-1, the injected source is the sum of two concentric de Vaucouleurs components with effective radii $R_{\rm e,1} =8\arcsec$ and $R_{\rm e,2} = 32\arcsec$. Both components have projected minor-to-major axis ratios $q=0.6$, and their combined profile is normalized to $\mu(20\arcsec)=24$ mag/arcsec$^2$. The compact component represents the main stellar content, while the broader component produces a diffuse envelope that extends far into the low surface brightness outskirts.

The double de Vaucouleurs profile (hereafter \texttt{ddV}) is motivated by the halo light decomposition studied in \citet{Ahad2023}. Using mock observations from the Hydrangea simulations, they modeled the central galaxy and surrounding intracluster and intragroup light with two de Vaucouleurs components describing the inner and outer light profiles. This is also observationally supported by \citet{Gonzalez2005} based on  measurements of the extended halos of nearby BCGs. 
We therefore use the \texttt{ddV} injection as an idealized massive galaxy/BCG+ICL-like system, mimicking a bright galaxy with an extended stellar envelope susceptible to cirrus contamination.

The Cutout-2 experiments aim for a different halo morphology. We inject a single exponential profile with $R_{\rm e}=20\arcsec$, central surface brightness $\mu_0=24.5$ mag/arcsec$^2$, and $q=0.65$. UDGs have a typical characteristic of S\'ersic index $n=1$ \citep[e.g.,][]{Koda2015,Mowla2017}. We therefore use the $n=1$ model as a UDG-like diffuse source, which is an experimental choice rather than placing a realistic UDG at the distance of Abell~2390.
The same exponential source is placed at three positions in Cutout-2, with the source fixed while altering the cirrus foreground. We illustrate with one injection in the main text, and present other injections in Appendix~\ref{app:injection}.

For each injection, we repeat the reconstruction from foreground initializations with independent phases, while holding the image, injected source, masks, covariance model, and other scientific inputs fixed. Radial profiles and cumulative flux curves are measured using the cirrus model subtracted image $\widetilde I$ (step C in Fig.~\ref{fig:workflow}). The profiles are measured with elliptical annuli following the geometry of the injected source, with other sources masked.  For the \texttt{ddV} profile, we adopt ellipse exclusion apertures with semi-major axis of $150\arcsec$ for conditioning initialization and $R_{\rm e,2}=32\arcsec$ for cirrus modeling, and measure the sky over $180\arcsec$--$220\arcsec$; for the exponential injections, the exclusion apertures $80\arcsec$ for initialization and $R_{\rm e}=20\arcsec$ for modeling, and the sky is measured over $100\arcsec$--$140\arcsec$. Uncertainties are estimated with a paired 12-sector jackknife, using the same angular sector for the source and the sky annulus. Uncertainties of multiple realizations are combined into the ensemble-averaged covariance. 

\begin{figure*}[!htbp]
\centering
  \resizebox{0.96\hsize}{!}{\includegraphics{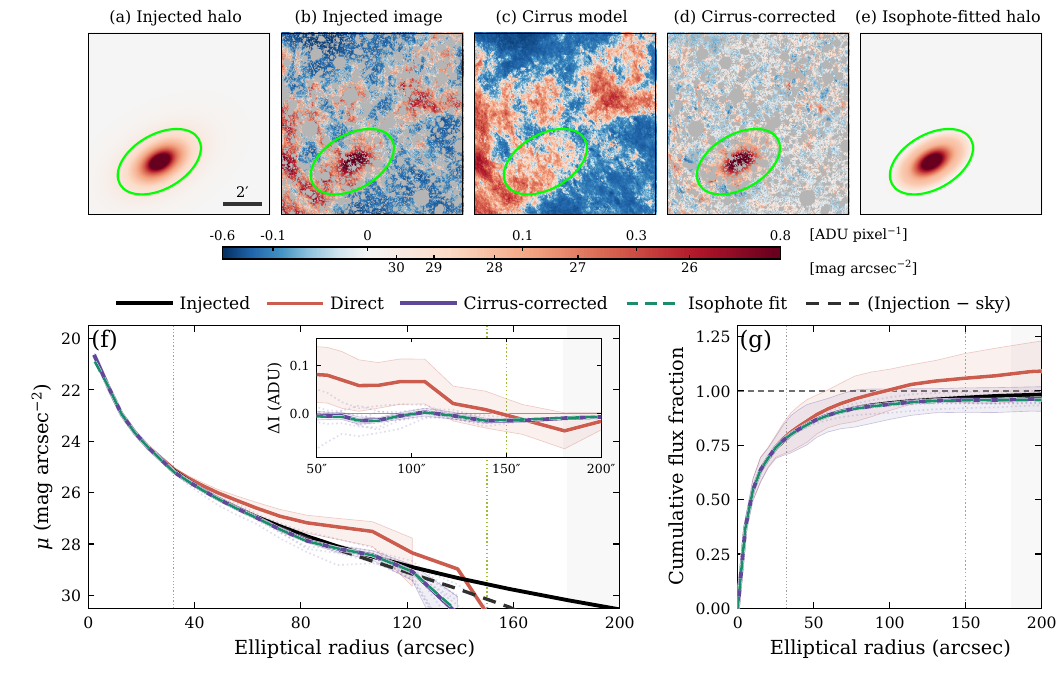}}
  \caption{\textbf{Recovery of the injected double de Vaucouleurs halo}. The upper panels show (a) the injected halo; (b) the source-masked Cutout-1 with injection; (c) one realization of cirrus reconstruction; (d) the ensemble-mean of cirrus-model subtracted image; and (e) the halo light recovery from the isophote fit. All image panels use the same asinh stretch. Panel (f) compares the directly measured, cirrus-corrected, and isophote model surface brightness profiles with the injected truth. The inset shows the linear intensity residual relative to the truth in the outskirts. Individual cirrus realizations are shown as dotted curves. The dashed black profile shows the sky-subtracted injected profile using the same sky annulus used for the measurements. Panel (g) shows the corresponding cumulative flux, normalized by the total injected flux with the same mask support. The core and target ellipse aperture radii are marked by gray and green dotted lines, respectively. The gray shaded regions indicate the sky annulus.}
\label{fig:cutout1_injection}
\end{figure*}

\begin{figure*}[!htbp]
\centering
  \resizebox{0.96\hsize}{!}{\includegraphics{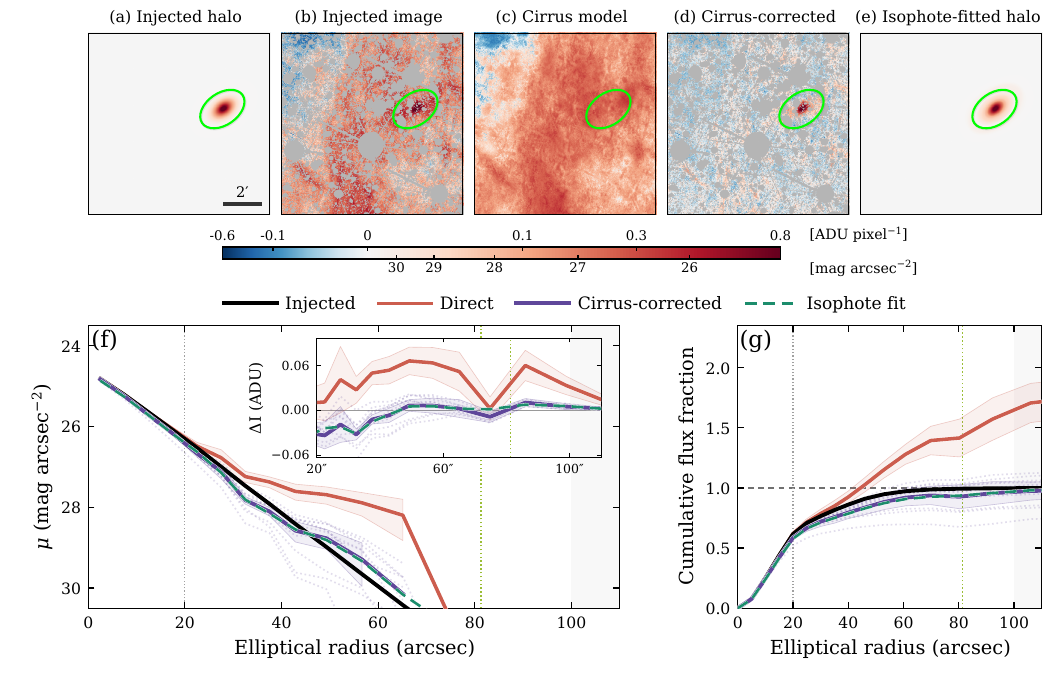}}
  \caption{\textbf{Recovery of the injected exponential halo}. The layout is the same as in Fig.~\ref{fig:cutout1_injection}, with injection in Cutout-2. Compared to the halo model in Fig.~\ref{fig:cutout1_injection}, this model represents a less extended, fainter, UDG-like diffuse halo. Cirrus contamination leaves a strong excess in the surface brightness and cumulative flux profiles.}
\label{fig:cutout2_injection_1}
\end{figure*}

\subsection{Halo light recovery across complex cirrus foregrounds}
\label{sec:halo_recovery}

Figures~\ref{fig:cutout1_injection} and \ref{fig:cutout2_injection_1} show the recovery for the two representative injections. In each figure, the image panels (a) to (e) show the injected model, the sourced-masked image with injection $I_{\rm obs}$, one realization of the reconstructed cirrus model $\widehat{I}_{\rm cir}$, the mean cirrus-subtracted image $\widetilde I$, and the smooth diffuse light model $\widehat I_{\rm diff}$ from the isophote fit. The cirrus-corrected panel shows the ensemble mean of the residual across ten realizations and is therefore smoother than individual reconstructions.

Fig.~\ref{fig:cutout1_injection}(f) compares the recovered and directly measured (i.e., with and without cirrus correction) surface brightness profiles with the input truth. The elliptical radius is defined as: 
$R_{\rm el}= \left[ x'^2+\left(y'/q\right)^2 \right] ^{1/2}$, where $x'$ and $y'$ align with the major and minor axes of the halo. 
The inset shows the linear intensity difference in analog-to-digital units (ADUs) relative to the truth at large radii. The lower right panel shows the enclosed flux as a function of radius, normalized by the total flux of the injected model with the same mask support. Measurements from individual foreground realizations are shown as dotted curves.

For the \texttt{ddV} experiment in Figure~\ref{fig:cutout1_injection}, the halo is placed near the edge of a bright cirrus patch where several filaments overlap, extending over a large angular size relative to the cutout field of view. Some halo light extends outside the target ellipse into the sky annulus. This is intentional, because in real data analysis it is practically difficult to define a clear boundary between halo and background in the presence of such strong contamination. To account for this effect in comparison, we applied a ``sky'' subtraction on the injection based on the measured halo light using the same sky annulus. The resulting profile is shown as the dashed black curve in Fig.~\ref{fig:cutout1_injection}(f).

Some small-scale cirrus structures remain visible in the residual image Fig.~\ref{fig:cutout1_injection}(d), while most of the cirrus light is removed. Also note that the image scale is stretched to emphasize faint structures. These residuals become further mitigated in the azimuthally averaged 1D profile of the halo light. The isophote fit hence further provides a smoother 2D representation of the diffuse halo. 

The cirrus-corrected profile well follows the injected halo profile over a wide dynamic range out to approximately $120\arcsec$, corresponding to $\mu \simeq 29$~mag/arcsec$^2$. At larger radii, the profile drops off, which may be caused by cirrus residuals or a slightly overestimated sky.
The cumulative flux is likewise well recovered over a wide radial range as shown in Fig.~\ref{fig:cutout1_injection}(g), whereas the direct measurement is biased high by $\simeq 9\%$ within the adopted elliptical aperture.
It is worth noting that one major improvement thanks to cirrus correction are the substantially reduced uncertainties at common radii, which effectively pushes the surface brightness depth that an averaged profile could achieve.

The $n=1$ injection in Fig.~\ref{fig:cutout2_injection_1} is placed on a prominent cirrus patch, making it a more challenging test. The sky estimate itself also includes strong cirrus distraction. The effects can be seen from its first flattened then sharply declined surface brightness profiles from direct measurement. The flux enclosed within $4\,R_{\rm e}$ has $\simeq 42\%$ excess above the truth, indicating a strong bias. After cirrus subtraction, much of this excess is removed. The profile follows that of the injected source over a wide radial range. The cumulative flux at $4\,R_{\rm e}$ is $\simeq 7\%$ lower than the truth, though much of the deficit is due to a 0.36~mag/arcsec$^2$ dip in the recovered profile near $1.6\, R_{\rm e}$. This halo model declines more rapidly, leading to a larger dispersion in individual realizations and a greater effect on the recovered outskirts. These can be seen in other injections presented in Appendix~\ref{app:injection} with the same halo model but at different positions.

These two experiments show that the method can recover diffuse halo light with different profile shapes and across different foreground patterns. The major improvements are a smaller systematic bias from coherent cirrus structure, reduced uncertainties, and more accurate sky estimates. The recovery is not perfect: some small-scale residuals remain; however, the residuals are usually lower in amplitude and lack large-scale coherence, leading to weaker imprints in the radial profile or cumulative flux. Improvements on their modeling will be explored in our future work.

\begin{figure*}[!htbp]
\centering
  \resizebox{0.96\hsize}{!}{\includegraphics{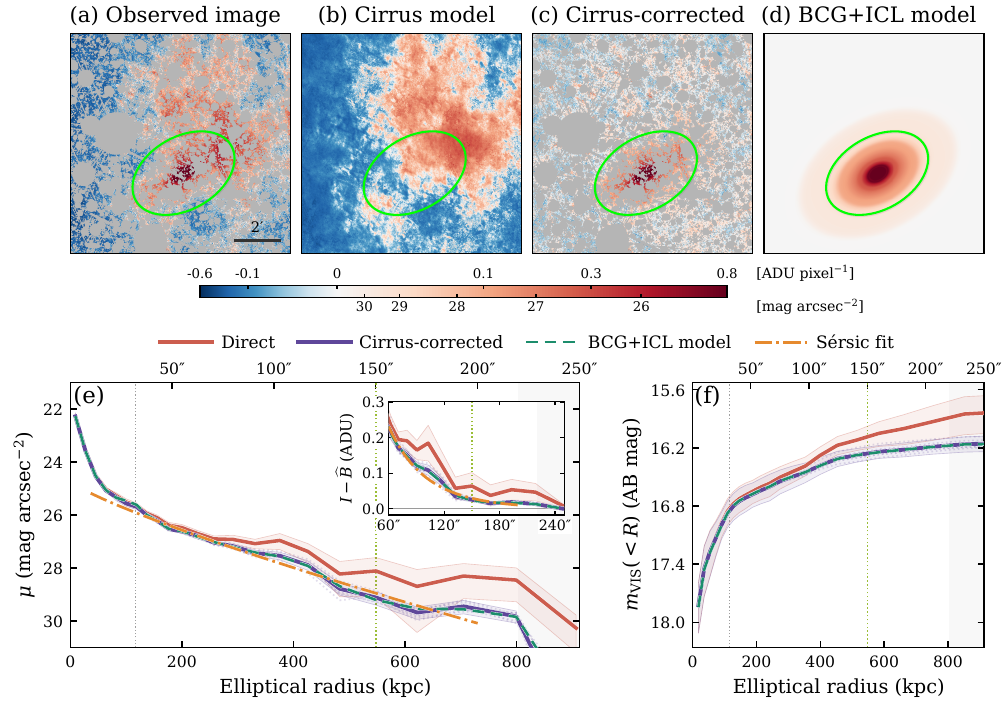}}
  \caption{\textbf{Application to the BCG+ICL in the central Abell~2390 field.}
  Panels (a)--(d) display the observed image, one cirrus model realization, ensemble-mean cirrus-corrected image, and smooth BCG+ICL isophote model, respectively. Panel~(e) compares the directly measured profile, the cirrus-subtracted surface brightness profile, and the smooth BCG+ICL model. The orange dash-dotted curve shows a single S\'ersic fit with $n=1.12$ to the corrected profile over $30\arcsec$--$200\arcsec$. The inset shows the intensity profiles in linear scale, after subtracting the sky estimate. Panel~(f) shows the enclosed VIS magnitude. Panels~(e) and (f) use elliptical annuli with geometry measured from the core part of the BCG. The cirrus correction leads to a steeper surface brightness profile and a more steady cumulative flux growth curve for the halo light measurement.}
\label{fig:A2390_bcg_icl}
\end{figure*}

\begin{figure*}[!htbp]
    \centering
    \includegraphics[width=\linewidth]{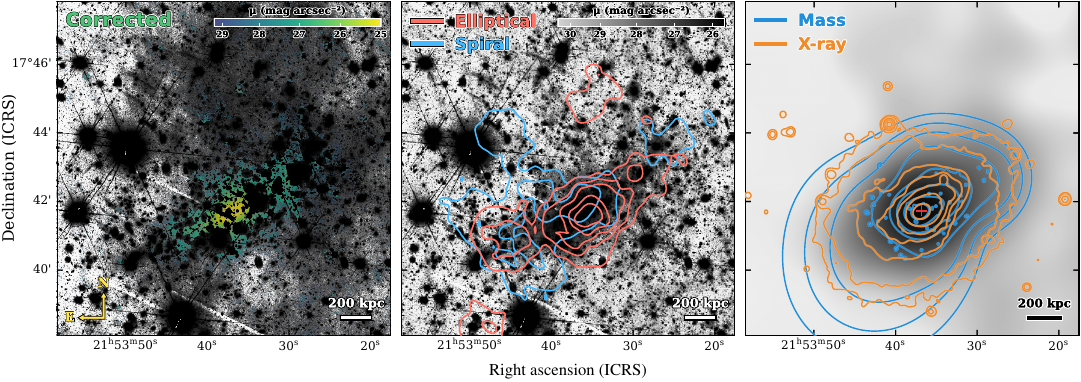}
  \caption{\textbf{Comparison of the cirrus-corrected diffuse emission with the member galaxy, total mass, and X-ray distributions in Abell~2390.}
  \textbf{Left:} The \textit{Euclid} VIS central field before cirrus subtraction (gray), overlaid with the cirrus-corrected diffuse emission.
  \textbf{Middle:} The VIS central field after cirrus correction,  with contours showing the spatial distributions of elliptical (red) and spiral (blue) cluster member galaxies. The diffuse emission spatially matches the elliptical galaxy distributions. \textbf{Right:} The cirrus-corrected BCG+ICL smoothed with a Gaussian kernel of $\sigma=36\arcsec$, overlaid with the contours of the total mass (blue) from lensing and hot-gas (orange) from X-ray. The mass distribution is from \cite{Diego2026}. The X-ray distribution is from \cite{Ellien2025}. The three components broadly match in their distributions.}
    \label{fig:A2390_overlay}
\end{figure*}

\section{Application to Abell 2390 ICL} \label{sec:application}

We now apply the foreground modeling to the BCG and ICL in the center of Abell~2390. Figure~\ref{fig:A2390_bcg_icl} shows a prominent cirrus complex crossing the northwestern part of the system, where its diffuse emission becomes comparable to the extended BCG+ICL at large radii. The cirrus model mostly follows this coherent foreground structure, and its subtraction reveals a substantially smoother distribution of diffuse cluster light. We measured the BCG+ICL from the cirrus-subtracted image, and performed an isophote fitting on this image. 

The effect of the foreground correction is most apparent in the outskirts. As shown in panel~(e), the direct profile shows a broad excess in the outer radii, whereas the cirrus-corrected profile declines more steeply with reduced measurement uncertainties. A single S\'ersic profile fitted over $30\arcsec$--$200\arcsec$ gives $n=1.12 \pm 0.15$, which is close to an exponential profile.

The cumulative photometry in panel~(f) shows the same behavior. We convert the cumulative flux to the total AB magnitude enclosed in the elliptical aperture $m_{\rm VIS}$. Without cirrus subtraction, the cumulative flux continues to increase at larger radii. The cirrus-subtracted measurement instead approaches a much flatter cumulative curve. 
This application therefore shows that the structured cirrus foreground can alter both the shape and integrated photometry of the observed BCG+ICL, and that modeling of the Galactic cirrus in the high-resolution image substantially increases the precision of the measurement.

Figure~\ref{fig:A2390_overlay} compares the recovered VIS BCG+ICL distribution with the member galaxy, total mass, and X-ray distributions. The left panel shows the cirrus-corrected diffuse emission overlaid on the original image. The middle panel shows the cirrus-corrected image overlaid with the spatial distributions of elliptical (red) and spiral (blue) cluster members.
These contours were constructed by \citet{Ellien2025} from visual classifications of 129 members within 1~Mpc of the BCG, selected based on spectroscopic redshifts from the NASA Extragalactic Database (NED) and membership probabilities from SDSS-redMaPPer \citep{Rykoff2014}; see \citet{Ellien2025} for details.

The recovered diffuse emission follows the elliptical population more closely than the spiral population, including the extensions southeast and northwest of the BCG. This agrees with the NISP analysis of \citet{Ellien2025}, who found that ellipticals trace the ICL, whereas spirals are more dispersed. They associated the ICL substructures with accreted galaxy groups, which suggests intragroup light formed before cluster infall being mixed into the ICL. This scenario is consistent with the orbits of emission line galaxies with detectable ionized gas offset near both substructures \citep{Liu2021}.

The right panel of Fig.~\ref{fig:A2390_overlay} shows the corrected diffuse emission in VIS with the total mass and X-ray distributions. For this comparison, we smoothed the source-masked cirrus-corrected VIS image with a Gaussian kernel of $\sigma=36\arcsec$ to account for the mask and emphasize the broad distribution. The mass and X-ray contours were constructed by \citet{Ellien2025} using strong+weak-lensing measurements \citep{Diego2026} and the Chandra Data Archive,\footnote{\url{https://cda.harvard.edu/chaser/}} respectively. The X-ray data is from Observation
4193 (PI: S. Allen).

The three components broadly follow the same spatial distributions with a northwest--southeast elongation.
Such morphological agreement is consistent with ICL being an effective tracer of the dark matter in galaxy clusters \citep[e.g.,][]{MontesTrujillo2019, Yoo2024Spatial5}.
This agreement, however, may not hold everywhere in the cluster. \citet{Ellien2025} found that these distributions differ for substructures and in the cluster outskirts. Such differences may provide clues to the cluster assembly history, reflecting the different dynamical responses of stars, dark matter, and hot gas. 
Here, we limit the comparison to a qualitative
assessment of the recovered VIS light morphology, since a detailed interpretation of the cluster dynamical
states is beyond the scope of this work. Overall, the agreement independently demonstrates the potential of our foreground correction method for studying the ICL.

\section{Implications for low surface brightness measurements} \label{sec:discussion}


\subsection{Sky estimate and foreground confusion for diffuse emission}
\label{sec:discussion_sky}

Galactic cirrus affects diffuse light photometry in a more complicated way than its direct contribution to the measured surface brightness. 
At the column density sensitivity where cirrus becomes detectable, the sky background can no longer be approximated as spatially uniform.\footnote{Even in sightlines with little Galactic dust, the astrophysical background contains spatial fluctuations from unresolved extragalactic sources. These include the measured anisotropies of the cosmic infrared background \citep{Viero2013} and the potential anisotropies of the cosmic optical background \citep{Zemcov2017, Postman2024}.}
There are two practical choices: subtraction of a global sky level estimated from relatively clean regions nearby\footnote{Finding such sky regions will be increasingly difficult as the next-generation deep imaging surveys reach fainter surface brightness limits.} leaves a significant systematic excess that can flatten the outer halo light profile. A local sky subtraction using a sky annulus may partly compensate for this bias, but it becomes sensitive to the cirrus structure across the source, the chosen sky annulus, as well as the background estimator. The resulting sky uncertainty becomes particularly important in the faint outskirts, where small sky offsets can produce artificial breaks, steepening, or upturns in galaxy surface brightness profiles  \citep[e.g.,][]{McDonald2011,Huang2018a,Li2022}.
Removing the spatially varying Galactic cirrus foreground therefore allows a more robust estimate of the local sky background.

Cirrus fluctuations are a source of astrophysical confusion noise. This effect is well known in FIR observations, where the structured emission from Galactic dust limits the detection and photometric accuracy of faint sources \citep{Gautier1992, Kiss2001, Jeong2005, MivilleDeschenes2007, Martin2010}. A similar effect arises in optical imaging when cirrus structures overlap the angular scales of stellar halos, tidal features, and ICL. These spatially correlated fluctuations do not average down as expected for Poisson statistics and can leave their imprints in radial profiles and aperture photometry. Increasing exposure time therefore cannot eliminate this source of confusion. Foreground modeling can reduce these structures, although residuals and uncertainties in the reconstruction remain part of the photometric error budget.

This is important for the effective surface brightness limit. Conventional limits are often calculated from background fluctuations over a fixed angular scale (e.g., \citealt{Trujillo2016, Keim2022, Cuillandre2025a}). The practical depth for detecting extended structures in a cirrus-riddled field, however, also depends on the overlap between cirrus and the source, as well as their surface brightness contrast.
Even for deep images reaching exceptionally faint nominal surface brightness limits, cirrus confusion may restrict any reliable detection of diffuse emission to substantially brighter levels. By removing much of this structured foreground, our approach can improve the effective surface brightness limit, which allows diffuse extragalactic emission to be measured closer to the depth that the observations can otherwise achieve.

\subsection{Limitations on reconstruction fidelity}
\label{sec:discussion_limitations}

The fidelity of the cirrus reconstruction depends on the information available to constrain its brightness and morphology. FIR observations provide a valuable prior to the broad dust distribution, although the correspondence between optical scattered light and thermal dust emission is imperfect at the desired resolution, as discussed in Sec.~\ref{sec:data_herschel}. In our implementation, a low-pass filter is used to suppress small-scale fluctuation from CIB anisotropies. A more complete separation of Galactic dust emission from CIB anisotropies, as demonstrated by \citet{Auclair2024} and \citet{Sinha2026}, could improve the foreground guide at small scales. The limited sky coverage of Herschel is another practical challenge. For wider sky areas, the WISE $12\,\mu$m all-sky dust map \citep{Meisner2014} or multiband optical color information may provide alternative guidance for the foreground, given appropriate calibration of their relations to optical cirrus surface brightness intensities.

The reliability of the reconstructed cirrus structure also depends on the pixel support of the data. Source masks prevent stars, galaxies, and instrumental artifacts from contaminating the multiscale statistics, but they also reduce the available information on intensity distributions, scale coupling, and directional statistics. As the masked area increases, fine structures within the masked region become less constrained. The reconstruction may then lose some small-scale contrast or fail to reproduce the orientation of individual filaments, even when its overall statistical properties agree with the observations. The modest orientation mismatch seen in our experiments (Figure~\ref{fig:app_orientation}) illustrates this uncertainty. 
Bright cirrus filaments and knots may also be
identified as extended sources by the detection
algorithm and masked. This preferentially
removes high-contrast cirrus structures and small-scale morphological information needed for their reconstruction. The loss of these constraints may explain some of the most prominent residuals around bright structures.

Mask design is therefore an important user choice. Aggressive masking protects the reconstruction from source contamination but leaves fewer pixels to constrain the cirrus. One improvement is to model and subtract sources including their extended PSF wings  \citep{Liu2022,Watkins2024,Eskandarlou2026}, although here we take advantage of the sharp \textit{Euclid} VIS PSF by simply masking them \citep{Cuillandre2025a}. Careful source modeling and masking followed by appropriate inpainting can potentially improve the reconstruction of the cirrus foreground that superposed on the target source.

\section{Conclusion} \label{sec:conclusion}

In this paper, we present a statistical approach to modeling and reconstructing optical Galactic cirrus in deep images. It combines the broad dust distribution traced by FIR emission with multiscale structural information from optical observations. 
The method synthesizes plausible cirrus structures by matching the target statistics, and simultaneously models the diffuse emission from a stellar halo behind the cirrus foreground. It is implemented in the \texttt{CirrusMachine}\footnote{\url{https://github.com/qliu4676/CirrusMachine}} package (Appendix~\ref{app:code}).

We perform injection--recovery experiments by injecting known halo light distributions into the cirrus-riddled \textit{Euclid} VIS image of the Abell~2390 cluster and assessing the photometry after cirrus correction.
We experiment with a double-de Vaucouleurs halo resembling the extended envelope of a massive galaxy or ICL, and a UDG-like exponential halo. The surface brightness profiles are well recovered out to the faint outskirts of the halos, whereas bias in the cumulative flux from aperture photometry is largely mitigated. 
We apply the method to the central Abell~2390 field by removing a prominent cirrus complex, which reveals a smoother BCG+ICL distribution with a steeper outer light profile. The recovered diffuse emission broadly follows the elliptical cluster members and the large-scale total mass and X-ray distributions, demonstrating the potential of foreground-corrected VIS imaging for studying ICL and cluster assembly.

These results highlight the importance of treating Galactic cirrus as a foreground systematic in low surface brightness measurements. Reducing the bias and confusion noise caused by cirrus fluctuations can facilitate the detection and photometry of diffuse emission with ongoing and forthcoming deep, wide-field imaging surveys.

\begin{acknowledgements}
Q.L. would like to thank Peter~Martin, Nina~Hatch, Marc-Antoine~Miville-Deschênes, Hua~Gao, Jean-Charles Cuillandre, Mathias~Urbano, Syeda~Lammim~Ahad, Mireia~Montes, Koen~Kuijken, David~Valls-Gabaud, and Pieter~van~Dokkum for helpful discussions. Q.L. is supported by the Oort Postdoctotal Fellowship.
This work has made use of the Early Release Observations (ERO) data from the {\it Euclid} mission of the European Space Agency (ESA), 2024, \url{https://doi.org/10.57780/esa-qmocze3}.
Herschel is an ESA space observatory with science instruments provided by European-led Principal Investigator consortia and with important participation from NASA.
\end{acknowledgements}

\bibliographystyle{aa} 
\bibliography{bibtex} 

@misc{EROcite,
author = "{Euclid Early Release Observations}",
howpublished = "\url{https://doi.org/10.57780/esa-qmocze3}",
year = 2024
}

@ARTICLE{EROLensData,
author = {{Atek}, H. and {Gavazzi}, R. and {Weaver}, J.R. and others},
	title = {Euclid: Early Release Observations â€“ A preview of the Euclid era through a galaxy cluster magnifying lens },
	DOI= "10.1051/0004-6361/202450776",
	url= "https://doi.org/10.1051/0004-6361/202450776",
	journal = {A\&A},
	year = 2025,
	volume = 697,
	pages = "A15",
}

@ARTICLE{Schrabback25,
       author = {{Schrabback}, T. and {Congedo}, G. and {Gavazzi}, R. and others},
       title = "{Euclid: Early Release Observations -- Weak gravitational lensing analysis of Abell 2390}",
      journal = {\aap},
         year = 2026,
        month = apr,
       volume = {708},
          eid = {A345},
        pages = {A345},
          doi = {10.1051/0004-6361/202556310},
archivePrefix = {arXiv},
       eprint = {2507.07629},
 primaryClass = {astro-ph.CO},
       adsurl = {https://ui.adsabs.harvard.edu/abs/2026A&A...708A.345S}
}

@ARTICLE{Liu2021,
       author = {{Liu}, Qing and {Yee}, H.~K.~C. and {Drissen}, Laurent and {Sivanandam}, Suresh and {Pintos-Castro}, Irene and {Alcorn}, Leo Y. and {Hsieh}, Bau-Ching and {Lin}, Lihwai and {Lin}, Yen-Ting and {Muzzin}, Adam and {Noble}, Allison and {Old}, Lyndsay},
        title = "{SITELLE H{\ensuremath{\alpha}} Imaging Spectroscopy of z {\ensuremath{\sim}} 0.25 Clusters: Emission-line Galaxy Detection and Ionized Gas Offset in Abell 2390 and Abell 2465}",
      journal = {\apj},
         year = 2021,
        month = feb,
       volume = {908},
       number = {2},
          eid = {228},
        pages = {228},
          doi = {10.3847/1538-4357/abd71e},
archivePrefix = {arXiv},
       eprint = {2101.01887},
 primaryClass = {astro-ph.GA},
       adsurl = {https://ui.adsabs.harvard.edu/abs/2021ApJ...908..228L}
}

@ARTICLE{laRosa2026,
       author = {{Marrero-de la Rosa}, Carlos and {Trujillo}, Ignacio and {Ruiz Cejudo}, Ignacio and {Montes}, Mireia and {Buitrago}, Fernando and {Golini}, Giulia and {Guerra Arencibia}, Sergio and {Asensio Ramos}, Andr{\'e}s and {Infante-Sainz}, Ra{\'u}l and {de Lorenzo-C{\'a}ceres}, Adriana and {M{\'e}ndez-Abreu}, Jairo and {Raji}, Samane and {Rom{\'a}n}, Javier and {S{\'a}nchez-Benavente}, Manuel and {Sharbaf}, Zahra},
        title = "{How large can galaxies be?: Ultra-deep imaging of IC 1101, the most extended known galaxy}",
      journal = {\aap},
         year = 2026,
        month = oct,
       volume = {714},
          eid = {A36},
        pages = {A36},
          doi = {10.1051/0004-6361/202660963},
       adsurl = {https://ui.adsabs.harvard.edu/abs/2026A&A...714A..36M}
}

@ARTICLE{Diego2026,
       author = {{Diego}, J.~M. and {Congedo}, G. and {Gavazzi}, R. and {Schrabback}, T. and {Atek}, H. and {Jain}, B. and {Weaver}, J.~R. and {Kang}, Y. and {Hartley}, W.~G. and {Mahler}, G. and {Okabe}, N. and {Golden-Marx}, J.~B. and {Meneghetti}, M. and {Palencia}, J.~M. and {Kluge}, M. and {Laureijs}, R. and {Saifollahi}, T. and {Schirmer}, M. and {Stone}, C. and {Jauzac}, M. and {Scott}, D. and {Altieri}, B. and {Amara}, A. and {Andreon}, S. and {Auricchio}, N. and {Baccigalupi}, C. and {Baldi}, M. and {Bardelli}, S. and {Battaglia}, P. and {Biviano}, A. and {Branchini}, E. and {Brescia}, M. and {Brinchmann}, J. and {Camera}, S. and {Ca{\~n}as-Herrera}, G. and {Candini}, G.~P. and {Capobianco}, V. and {Carbone}, C. and {Cardone}, V.~F. and {Carretero}, J. and {Casas}, S. and {Castellano}, M. and {Castignani}, G. and {Cavuoti}, S. and {Chambers}, K.~C. and {Cimatti}, A. and {Colodro-Conde}, C. and {Conselice}, C.~J. and {Conversi}, L. and {Copin}, Y. and {Courbin}, F. and {Courtois}, H.~M. and {Cropper}, M. and {Cuillandre}, J.-C. and {Da Silva}, A. and {Degaudenzi}, H. and {De Lucia}, G. and {Dole}, H. and {Douspis}, M. and {Dubath}, F. and {Dupac}, X. and {Dusini}, S. and {Escoffier}, S. and {Farina}, M. and {Farrens}, S. and {Faustini}, F. and {Ferriol}, S. and {Finelli}, F. and {Fosalba}, P. and {Fourmanoit}, N. and {Frailis}, M. and {Franceschi}, E. and {Fumana}, M. and {Galeotta}, S. and {George}, K. and {Gillis}, B. and {Giocoli}, C. and {Gracia-Carpio}, J. and {Grazian}, A. and {Grupp}, F. and {Guzzo}, L. and {Haugan}, S.~V.~H. and {Hoar}, J. and {Holmes}, W. and {Hook}, I.~M. and {Hormuth}, F. and {Hornstrup}, A. and {Hudelot}, P. and {Jahnke}, K. and {Jhabvala}, M. and {Joachimi}, B. and {Keih{\"a}nen}, E. and {Kermiche}, S. and {Kilbinger}, M. and {Kubik}, B. and {Kuijken}, K. and {K{\"u}mmel}, M. and {Kunz}, M. and {Kurki-Suonio}, H. and {Le Brun}, A.~M.~C. and {Le Mignant}, D. and {Ligori}, S. and {Lilje}, P.~B. and {Lindholm}, V. and {Lloro}, I. and {Mainetti}, G. and {Maino}, D. and {Maiorano}, E. and {Mansutti}, O. and {Marggraf}, O. and {Martinelli}, M. and {Martinet}, N. and {Marulli}, F. and {Massey}, R.~J. and {Medinaceli}, E. and {Mei}, S. and {Melchior}, M. and {Mellier}, Y. and {Merlin}, E. and {Meylan}, G. and {Mohr}, J.~J. and {Mora}, A. and {Moresco}, M. and {Moscardini}, L. and {Munari}, E. and {Nakajima}, R. and {Neissner}, C. and {Nichol}, R.~C. and {Niemi}, S.-M. and {Padilla}, C. and {Paltani}, S. and {Pasian}, F. and {Pedersen}, K. and {Percival}, W.~J. and {Pettorino}, V. and {Pires}, S. and {Polenta}, G. and {Poncet}, M. and {Popa}, L.~A. and {Pozzetti}, L. and {Raison}, F. and {Renzi}, A. and {Rhodes}, J. and {Riccio}, G. and {Romelli}, E. and {Roncarelli}, M. and {Saglia}, R. and {Sakr}, Z. and {Sapone}, D. and {Sartoris}, B. and {Schneider}, P. and {Secroun}, A. and {Seidel}, G. and {Seiffert}, M. and {Serrano}, S. and {Simon}, P. and {Sirignano}, C. and {Sirri}, G. and {Stanco}, L. and {Steinwagner}, J. and {Tallada-Cresp{\'\i}}, P. and {Taylor}, A.~N. and {Tereno}, I. and {Tessore}, N. and {Toft}, S. and {Toledo-Moreo}, R. and {Torradeflot}, F. and {Tutusaus}, I. and {Valenziano}, L. and {Valiviita}, J. and {Vassallo}, T. and {Verdoes Kleijn}, G. and {Veropalumbo}, A. and {Wang}, Y. and {Weller}, J. and {Zamorani}, G. and {Zerbi}, F.~M. and {Zucca}, E. and {Bolzonella}, M. and {Burigana}, C. and {Gabarra}, L. and {Mart{\'\i}n-Fleitas}, J. and {Matthew}, S. and {Scottez}, V. and {Sereno}, M. and {Viel}, M.},
        title = "{Euclid: Early Release Observations -- A combined strong and weak lensing solution for Abell 2390 beyond its virial radius}",
      journal = {\aap},
         year = 2026,
        month = feb,
       volume = {706},
          eid = {A83},
        pages = {A83},
          doi = {10.1051/0004-6361/202556372},
archivePrefix = {arXiv},
       eprint = {2507.08545},
 primaryClass = {astro-ph.CO},
       adsurl = {https://ui.adsabs.harvard.edu/abs/2026A&A...706A..83D}
}

@ARTICLE{MontesTrujillo2019,
       author = {{Montes}, Mireia and {Trujillo}, Ignacio},
        title = "{Intracluster light: a luminous tracer for dark matter in clusters of galaxies}",
      journal = {\mnras},
         year = 2019,
        month = jan,
       volume = {482},
       number = {2},
        pages = {2838-2851},
          doi = {10.1093/mnras/sty2858},
archivePrefix = {arXiv},
       eprint = {1807.11488},
 primaryClass = {astro-ph.GA},
       adsurl = {https://ui.adsabs.harvard.edu/abs/2019MNRAS.482.2838M}
}

@ARTICLE{Rykoff2014,
       author = {{Rykoff}, E.~S. and {Rozo}, E. and {Busha}, M.~T. and {Cunha}, C.~E. and {Finoguenov}, A. and {Evrard}, A. and {Hao}, J. and {Koester}, B.~P. and {Leauthaud}, A. and {Nord}, B. and {Pierre}, M. and {Reddick}, R. and {Sadibekova}, T. and {Sheldon}, E.~S. and {Wechsler}, R.~H.},
        title = "{redMaPPer. I. Algorithm and SDSS DR8 Catalog}",
      journal = {\apj},
         year = 2014,
        month = apr,
       volume = {785},
       number = {2},
          eid = {104},
        pages = {104},
          doi = {10.1088/0004-637X/785/2/104},
archivePrefix = {arXiv},
       eprint = {1303.3562},
 primaryClass = {astro-ph.CO},
       adsurl = {https://ui.adsabs.harvard.edu/abs/2014ApJ...785..104R}
}

@ARTICLE{BaloghMorris2000,
       author = {{Balogh}, Michael L. and {Morris}, Simon L.},
        title = "{H{\ensuremath{\alpha}} photometry of Abell 2390}",
      journal = {\mnras},
         year = 2000,
        month = nov,
       volume = {318},
       number = {3},
        pages = {703-714},
          doi = {10.1046/j.1365-8711.2000.03826.x},
archivePrefix = {arXiv},
       eprint = {astro-ph/0007111},
 primaryClass = {astro-ph},
       adsurl = {https://ui.adsabs.harvard.edu/abs/2000MNRAS.318..703B}
}

@ARTICLE{Abraham1996,
       author = {{Abraham}, R.~G. and {Smecker-Hane}, Tammy A. and {Hutchings}, J.~B. and {Carlberg}, R.~G. and {Yee}, H.~K.~C. and {Ellingson}, Erica and {Morris}, Simon and {Oke}, J.~B. and {Rigler}, Michael},
        title = "{Galaxy Evolution in Abell 2390}",
      journal = {\apj},
         year = 1996,
        month = nov,
       volume = {471},
        pages = {694},
          doi = {10.1086/177999},
archivePrefix = {arXiv},
       eprint = {astro-ph/9605144},
 primaryClass = {astro-ph},
       adsurl = {https://ui.adsabs.harvard.edu/abs/1996ApJ...471..694A}
}

@ARTICLE{Yee1996,
       author = {{Yee}, H.~K.~C. and {Ellingson}, E. and {Abraham}, R.~G. and {Gravel}, Pierre and {Carlberg}, R.~G. and {Smecker-Hane}, Tammy A. and {Schade}, David and {Rigler}, Michael},
        title = "{The CNOC Cluster Redshift Survey Catalogs. II. Abell 2390}",
      journal = {\apjs},
         year = 1996,
        month = feb,
       volume = {102},
        pages = {289},
          doi = {10.1086/192260},
       adsurl = {https://ui.adsabs.harvard.edu/abs/1996ApJS..102..289Y}
}

@article{Abraham2014,
  author = {Abraham, Roberto G. and van Dokkum, Pieter},
  title = {{Ultra Low Surface Brightness Imaging with the Dragonfly Telephoto Array}},
  year = {2014},
  eprint = {1401.5473},
  archivePrefix = {arXiv},
  doi = {10.1086/674875},
  journal = {Publications of the Astronomical Society of the Pacific},
  volume = {126},
  pages = {55},
  url = {https://doi.org/10.1086/674875}
}

@article{Ahad2023,
  author = {Ahad, Syeda Lammim and Bah{\'{e}}, Yannick M. and Hoekstra, Henk},
  title = {{How to Interpret Measurements of Diffuse Light in Stacked Observations of Groups and Clusters of Galaxies}},
  year = {2023},
  eprint = {2210.14249},
  archivePrefix = {arXiv},
  doi = {10.1093/mnras/stac3357},
  journal = {Monthly Notices of the Royal Astronomical Society},
  volume = {518},
  pages = {3685--3701},
  url = {https://doi.org/10.1093/mnras/stac3357}
}

@article{Allys2019,
  author = {Allys, E. and Levrier, F. and Zhang, S. and Colling, C. and Blancard, B. Regaldo-Saint and Boulanger, F. and Hennebelle, P. and Mallat, S.},
  title = {{The RWST, a comprehensive statistical description of the non-Gaussian structures in the ISM}},
  year = {2019},
  eprint = {1905.01372},
  archivePrefix = {arXiv},
  doi = {10.1051/0004-6361/201834975},
  journal = {Astronomy \& Astrophysics},
  volume = {629},
  pages = {A115},
  url = {https://doi.org/10.1051/0004-6361/201834975}
}

@article{Auclair2024,
  author = {Auclair, C. and Allys, E. and Boulanger, F. and others},
  title = {{Separation of dust emission from the Cosmic Infrared Background in Herschel observations with Wavelet Phase Harmonics}},
  journal = {Astronomy \& Astrophysics},
  year = {2024},
  volume = {681},
  eid = {A1},
  doi = {10.1051/0004-6361/202346814},
  eprint = {2305.14419},
  archivePrefix = {arXiv},
  pages = {A1},
  url = {https://doi.org/10.1051/0004-6361/202346814}
}

@article{Bilek2020,
  author = {B{\'{i}}lek, Michal and Duc, Pierre-Alain and Cuillandre, Jean-Charles and others},
  title = {{Census and classification of low-surface-brightness structures in nearby early-type galaxies from the MATLAS survey}},
  year = {2020},
  eprint = {2007.13772},
  archivePrefix = {arXiv},
  doi = {10.1093/mnras/staa2248},
  journal = {Monthly Notices of the Royal Astronomical Society},
  volume = {498},
  pages = {2138--2166},
  url = {https://doi.org/10.1093/mnras/staa2248}
}

@inproceedings{adam,
  author = {Kingma, Diederik P. and Ba, Jimmy},
  booktitle = {ICLR (Poster)},
  crossref = {conf/iclr/2015},
  editor = {Bengio, Yoshua and LeCun, Yann},
  ee = {https://doi.org/10.48550/arXiv.1412.6980},
  title = {Adam: A Method for Stochastic Optimization.},
  url = {http://dblp.uni-trier.de/db/conf/iclr/iclr2015.html#KingmaB14},
  year = 2015
}

@software{Bradley2016,
       author = {{Bradley}, Larry and {Sipocz}, Brigitta and {Robitaille}, Thomas and {Tollerud}, Erik and {Deil}, Christoph and {Vin{\'\i}cius}, Z{\`e} and {Barbary}, Kyle and {G{\"u}nther}, Hans Moritz and {Bostroem}, Azalee and {Droettboom}, Michael and {Bray}, Erik and {Bratholm}, Lars Andersen and {Pickering}, T.~E. and {Craig}, Matt and {Pascual}, Sergio and {Greco}, Johnny and {Donath}, Axel and {Kerzendorf}, Wolfgang and {Littlefair}, Stuart and {Barentsen}, Geert and {D'Eugenio}, Francesco and {Weaver}, Benjamin Alan},
        title = "{Photutils: Photometry tools}",
 howpublished = {Astrophysics Source Code Library, record ascl:1609.011},
         year = 2016,
        month = sep,
          eid = {ascl:1609.011},
archivePrefix = {ascl},
       eprint = {1609.011},
       adsurl = {https://ui.adsabs.harvard.edu/abs/2016ascl.soft09011B}
}

@article{Bullock2005,
  author = {Bullock, James S. and Johnston, Kathryn V.},
  title = {{Tracing Galaxy Formation with Stellar Halos I: Methods}},
  year = {2005},
  eprint = {astro-ph/0506467},
  archivePrefix = {arXiv},
  doi = {10.1086/497422},
  journal = {The Astrophysical Journal},
  volume = {635},
  pages = {931--949},
  url = {https://doi.org/10.1086/497422}
}

@ARTICLE{Burkhart2020,
       author = {{Burkhart}, B. and {Appel}, S.~M. and {Bialy}, S. and {Cho}, J. and {Christensen}, A.~J. and {Collins}, D. and {Federrath}, C. and {Fielding}, D.~B. and {Finkbeiner}, D. and {Hill}, A.~S. and {Ib{\'a}{\~n}ez-Mej{\'\i}a}, J.~C. and {Krumholz}, M.~R. and {Lazarian}, A. and {Li}, M. and {Mocz}, P. and {Mac Low}, M.-M. and {Naiman}, J. and {Portillo}, S.~K.~N. and {Shane}, B. and {Slepian}, Z. and {Yuan}, Y.},
        title = "{The Catalogue for Astrophysical Turbulence Simulations (CATS)}",
      journal = {\apj},
         year = 2020,
        month = dec,
       volume = {905},
       number = {1},
          eid = {14},
        pages = {14},
          doi = {10.3847/1538-4357/abc484},
archivePrefix = {arXiv},
       eprint = {2010.11227},
 primaryClass = {astro-ph.GA},
       adsurl = {https://ui.adsabs.harvard.edu/abs/2020ApJ...905...14B}
}

@article{ChengMenard2020,
  author = {Cheng, Sihao and Ting, Yuan-Sen and M{\'{e}}nard, Brice and Bruna, Joan},
  title = {{A new approach to observational cosmology using the scattering transform}},
  year = {2020},
  journal = {Monthly Notices of the Royal Astronomical Society},
  volume = {499},
  number = {4},
  pages = {5902--5914},
  doi = {10.1093/mnras/staa3165},
  eprint = {2006.08561},
  archivePrefix = {arXiv},
  primaryClass = {astro-ph.CO},
  url = {https://doi.org/10.1093/mnras/staa3165}
}

@article{Clark2014,
  author = {{Clark}, S. E. and {Peek}, J. E. G. and {Putman}, M. E.},
  title = {{Magnetically Aligned H I Fibers and the Rolling Hough Transform}},
  journal = {The Astrophysical Journal},
  year = {2014},
  volume = {789},
  eid = {82},
  doi = {10.1088/0004-637X/789/1/82},
  eprint = {1312.1338},
  archivePrefix = {arXiv},
  pages = {82},
  url = {https://doi.org/10.1088/0004-637X/789/1/82}
}

@article{Cortese2010,
  author = {{Cortese}, L. and {Bendo}, G. J. and {Isaak}, K. G. and {Davies}, J. I. and {Kent}, B. R.},
  title = {{Diffuse far-infrared and ultraviolet emission in the NGC 4435/4438 system: tidal stream or Galactic cirrus?}},
  journal = {Monthly Notices of the Royal Astronomical Society: Letters},
  year = {2010},
  volume = {403},
  pages = {L26--L30},
  doi = {10.1111/j.1745-3933.2009.00808.x},
  eprint = {1001.0980},
  archivePrefix = {arXiv},
  url = {https://doi.org/10.1111/j.1745-3933.2009.00808.x}
}

@article{Cuillandre2025a,
  author = {Cuillandre, J.-C. and Bertin, E. and Bolzonella, M. and others},
  title = {{Euclid: Early Release Observations. Programme overview and pipeline for compact- and diffuse-emission photometry}},
  journal = {Astronomy \& Astrophysics},
  year = {2025},
  volume = {697},
  eid = {A6},
  doi = {10.1051/0004-6361/202450803},
  eprint = {2405.13496},
  archivePrefix = {arXiv},
  pages = {A6},
  url = {https://doi.org/10.1051/0004-6361/202450803}
}

@article{DSouza2018,
  author = {D'Souza, Richard and Bell, Eric},
  title = {{The masses and metallicities of stellar haloes reflect galactic merger histories}},
  year = {2018},
  eprint = {1705.08442},
  archivePrefix = {arXiv},
  doi = {10.1093/mnras/stx3081},
  journal = {Monthly Notices of the Royal Astronomical Society},
  volume = {474},
  pages = {5300},
  url = {https://doi.org/10.1093/mnras/stx3081}
}

@article{Duc2015,
  author = {Duc, Pierre-Alain and Cuillandre, Jean-Charles and Karabal, Emin and others},
  title = {{The ATLAS$^{3D}$ project -- XXIX. The new look of early-type galaxies and surrounding fields disclosed by extremely deep optical images}},
  year = {2015},
  journal = {Monthly Notices of the Royal Astronomical Society},
  volume = {446},
  number = {1},
  pages = {120--143},
  doi = {10.1093/mnras/stu2019},
  eprint = {1410.0981},
  archivePrefix = {arXiv},
  adsurl = {https://ui.adsabs.harvard.edu/abs/2015MNRAS.446..120D/abstract},
  url = {https://doi.org/10.1093/mnras/stu2019}
}

@article{ElBadry2016,
  author = {El-Badry, Kareem and Wetzel, Andrew R. and Geha, Marla and Hopkins, Philip F. and Kere{\v{s}}, Du{\v{s}}an and Chan, T. K. and Faucher-Gigu{\`{e}}re, Claude-Andr{\'{e}}},
  title = {{Breathing FIRE: How Stellar Feedback Drives Radial Migration, Rapid Size Fluctuations, and Population Gradients in Low-Mass Galaxies}},
  year = {2016},
  eprint = {1512.01235},
  archivePrefix = {arXiv},
  doi = {10.3847/0004-637X/820/2/131},
  journal = {The Astrophysical Journal},
  volume = {820},
  pages = {131},
  url = {https://doi.org/10.3847/0004-637X/820/2/131}
}

@ARTICLE{Ellien2021,
       author = {{Ellien}, A. and {Slezak}, E. and {Martinet}, N. and {Durret}, F. and {Adami}, C. and {Gavazzi}, R. and {Raba{\c{c}}a}, C.~R. and {Da Rocha}, C. and {Epit{\'a}cio Pereira}, D.~N.},
        title = "{DAWIS: a detection algorithm with wavelets for intracluster light studies}",
      journal = {\aap},
         year = 2021,
        month = may,
       volume = {649},
          eid = {A38},
        pages = {A38},
          doi = {10.1051/0004-6361/202038419},
archivePrefix = {arXiv},
       eprint = {2101.03835},
 primaryClass = {astro-ph.GA},
       adsurl = {https://ui.adsabs.harvard.edu/abs/2021A&A...649A..38E}
}

@article{Ellien2025,
  author = {Ellien, A. and Montes, M. and Ahad, S. L. and others},
  title = {{Euclid: Early Release Observations. The intracluster light of Abell 2390}},
  journal = {Astronomy \& Astrophysics},
  year = {2025},
  volume = {698},
  eid = {A134},
  doi = {10.1051/0004-6361/202554460},
  pages = {A134},
  url = {https://doi.org/10.1051/0004-6361/202554460}
}

@article{Eskandarlou2026,
  author = {Eskandarlou, Sepideh and Akhlaghi, Mohammad and Knapen, Johan H. and others},
  title = {{J-PLUS: Turning Off the Bright Stars}},
  year = {2026},
  eprint = {2510.12940},
  archivePrefix = {arXiv},
  doi = {10.1051/0004-6361/202556547},
  journal = {Astronomy \& Astrophysics},
  volume = {705},
  pages = {A67},
  url = {https://doi.org/10.1051/0004-6361/202556547}
}

@article{FliriTrujillo2016,
  author = {{Fliri}, J. and {Trujillo}, I.},
  title = {{The {IAC Stripe 82 Legacy Project}: a wide-area survey for faint surface brightness astronomy}},
  journal = {Monthly Notices of the Royal Astronomical Society},
  year = {2016},
  volume = {456},
  pages = {1359--1373},
  doi = {10.1093/mnras/stv2686},
  eprint = {1603.04474},
  archivePrefix = {arXiv},
  url = {https://doi.org/10.1093/mnras/stv2686}
}

@article{Gautier1992,
  author = {Gautier, III, T. N. and Boulanger, F. and P{\'e}rault, M. and Puget, J.-L.},
  title = {{A Calculation of Confusion Noise Due to Infrared Cirrus}},
  journal = {The Astronomical Journal},
  year = {1992},
  volume = {103},
  pages = {1313--1324},
  doi = {10.1086/116144},
  url = {https://doi.org/10.1086/116144}
}

@article{Gilhuly2022,
  author = {Gilhuly, Colleen and Merritt, Allison and Abraham, Roberto and Danieli, Shany and Lokhorst, Deborah and Liu, Qing and van Dokkum, Pieter and Conroy, Charlie and Greco, Johnny},
  title = {{Stellar halos from The Dragonfly Edge-on Galaxies Survey}},
  year = {2022},
  eprint = {2204.06596},
  archivePrefix = {arXiv},
  doi = {10.3847/1538-4357/ac6750},
  journal = {The Astrophysical Journal},
  volume = {932},
  pages = {44},
  url = {https://doi.org/10.3847/1538-4357/ac6750}
}

@article{Gonzalez2005,
  author = {Gonzalez, Anthony H. and Zabludoff, Ann I. and Zaritsky, Dennis},
  title = {{Intracluster Light in Nearby Galaxy Clusters: Relationship to the Halos of Brightest Cluster Galaxies}},
  doi = {10.1086/425896},
  year = {2005},
  journal = {The Astrophysical Journal},
  volume = {618},
  pages = {195--213},
  url = {https://doi.org/10.1086/425896}
}

@article{Griffin2010,
  author = {Griffin, M. J. and Abergel, A. and Abreu, A. and others},
  title = {{The Herschel-SPIRE instrument and its in-flight performance}},
  year = {2010},
  eprint = {1005.5123},
  archivePrefix = {arXiv},
  doi = {10.1051/0004-6361/201014519},
  journal = {Astronomy \& Astrophysics},
  volume = {518},
  pages = {L3},
  url = {https://doi.org/10.1051/0004-6361/201014519}
}

@article{Guhathakurta1989,
  author = {{Guhathakurta}, P. and {Tyson}, J.~A.},
  title = {{{Optical Characteristics of Galactic 100 Micron Cirrus}}},
  journal = {The Astrophysical Journal},
  year = {1989},
  volume = {346},
  pages = {773},
  doi = {10.1086/168058},
  adsurl = {https://ui.adsabs.harvard.edu/abs/1989ApJ...346..773G},
  url = {https://doi.org/10.1086/168058}
}

@article{Heesters2025,
  author = {Heesters, Nick and Chemaly, David and M{\"{u}}ller, Oliver and others},
  title = {{Galaxies OBserved as Low-luminosity Identified Nebulae (GOBLIN): a catalog of 43,000 high-probability dwarf galaxy candidates in the UNIONS survey}},
  year = {2025},
  eprint = {2505.18307},
  archivePrefix = {arXiv},
  doi = {10.1051/0004-6361/202554501},
  journal = {Astronomy \& Astrophysics},
  volume = {699},
  pages = {A232},
  url = {https://doi.org/10.1051/0004-6361/202554501}
}

@ARTICLE{HoffmanRibak1991,
       author = {{Hoffman}, Yehuda and {Ribak}, Erez},
        title = "{Constrained Realizations of Gaussian Fields: A Simple Algorithm}",
      journal = {\apjl},
         year = 1991,
        month = oct,
       volume = {380},
        pages = {L5},
          doi = {10.1086/186160},
       adsurl = {https://ui.adsabs.harvard.edu/abs/1991ApJ...380L...5H}
}

@article{Huang2018a,
  author = {Huang, Song and Leauthaud, Alexie and Greene, Jenny and Bundy, Kevin and Lin, Yen-Ting and Tanaka, Masayuki and Miyazaki, Satoshi and Komiyama, Yutaka},
  title = {{Individual Stellar Halos of Massive Galaxies Measured to 100 kpc at $0.3<z<0.5$ using Hyper Suprime-Cam}},
  year = {2018},
  eprint = {1707.01904},
  archivePrefix = {arXiv},
  doi = {10.1093/mnras/stx3200},
  journal = {Monthly Notices of the Royal Astronomical Society},
  volume = {475},
  pages = {3348--3368},
  url = {https://doi.org/10.1093/mnras/stx3200}
}

@article{Ienaka2013,
  author = {Ienaka, N. and Kawara, K. and Matsuoka, Y. and Sameshima, H. and Oyabu, S. and Tsujimoto, T. and Peterson, B. A.},
  title = {{Diffuse Galactic Light in the Field of the Translucent High Galactic Latitude Cloud MBM32}},
  journal = {The Astrophysical Journal},
  year = {2013},
  volume = {767},
  eid = {80},
  doi = {10.1088/0004-637X/767/1/80},
  eprint = {1303.0938},
  archivePrefix = {arXiv},
  pages = {80},
  url = {https://doi.org/10.1088/0004-637X/767/1/80}
}

@article{InfanteSainz2020,
  author = {Infante-Sainz, Ra{\'{u}}l and Trujillo, Ignacio and Rom{\'{a}}n, Javier},
  title = {{The Sloan Digital Sky Survey extended point spread functions}},
  year = {2020},
  eprint = {1911.01430},
  archivePrefix = {arXiv},
  doi = {10.1093/mnras/stz3111},
  journal = {Monthly Notices of the Royal Astronomical Society},
  volume = {491},
  pages = {5317--5329},
  url = {https://doi.org/10.1093/mnras/stz3111}
}

@article{InoueInutsuka2016,
  author = {Inoue, Tsuyoshi and Inutsuka, Shu-ichiro},
  title = {{Formation of HI Clouds in Shock-compressed Interstellar Medium: Physical Origin of Angular Correlation Between Filamentary Structure and Magnetic Field}},
  year = {2016},
  eprint = {1608.02601},
  archivePrefix = {arXiv},
  doi = {10.3847/0004-637X/833/1/10},
  journal = {The Astrophysical Journal},
  volume = {833},
  pages = {10},
  url = {https://doi.org/10.3847/0004-637X/833/1/10}
}

@article{Jeong2005,
  author = {Jeong, Woong-Seob and Lee, Hyung Mok and Pak, Soojong and Nakagawa, Takao and Kwon, Suk Minn and Pearson, Chris P. and White, Glenn J.},
  title = {{Far-infrared detection limits -- I. Sky confusion due to Galactic cirrus}},
  doi = {10.1111/j.1365-2966.2005.08627.x},
  year = {2005},
  journal = {Monthly Notices of the Royal Astronomical Society},
  volume = {357},
  pages = {535--547},
  url = {https://doi.org/10.1111/j.1365-2966.2005.08627.x}
}

@misc{Kaboud2026,
  author = {Kaboud, Ouldouz and Trujillo, Ignacio and Cejudo, Ignacio Ruiz and others},
  title = {{LIGHTS. Limitations of Far-Infrared Tracers for Galactic Cirrus Mitigation in the Era of Ultra-Deep Imaging}},
  year = {2026},
  eprint = {2610.00463},
  archivePrefix = {arXiv},
  howpublished = {arXiv e-prints},
  note = {Accepted for publication in Astronomy \& Astrophysics},
  url = {https://arxiv.org/abs/2610.00463}
}

@article{KadoFong2020,
  author = {Kado-Fong, Erin and Greene, Jenny E. and Huang, Song and Beaton, Rachael and Goulding, Andy D. and Komiyama, Yutaka},
  title = {{Tracing the Intrinsic Shapes of Dwarf Galaxies out to Four Effective Radii: Clues to Low-Mass Stellar Halo Formation}},
  year = {2020},
  eprint = {2007.10349},
  archivePrefix = {arXiv},
  journal = {The Astrophysical Journal},
  volume = {900},
  pages = {163},
  doi = {10.3847/1538-4357/abacc2},
  url = {https://doi.org/10.3847/1538-4357/abacc2}
}

@article{Keim2022,
  author = {Keim, Michael A. and van Dokkum, Pieter and Danieli, Shany and Lokhorst, Deborah and Li, Jiaxuan and Shen, Zili and Abraham, Roberto and Chen, Seery and Gilhuly, Colleen and Liu, Qing and Merritt, Allison and Miller, Tim B. and Pasha, Imad and Polzin, Ava},
  title = {{Tidal Distortions in NGC1052-DF2 and NGC1052-DF4: Independent Evidence for a Lack of Dark Matter}},
  doi = {10.3847/1538-4357/ac7dab},
  year = {2022},
  journal = {The Astrophysical Journal},
  volume = {935},
  pages = {160},
  url = {https://doi.org/10.3847/1538-4357/ac7dab}
}

@article{Kiss2001,
  author = {Kiss, Cs. and Abraham, P. and Klaas, U. and Juvela, M. and Lemke, D.},
  title = {{Sky Confusion Noise in the Far-Infrared: Cirrus, Galaxies and the Cosmic Far-Infrared Background}},
  year = {2001},
  eprint = {astro-ph/0110143},
  archivePrefix = {arXiv},
  doi = {10.1051/0004-6361:20011394},
  journal = {Astronomy \& Astrophysics},
  volume = {379},
  pages = {1161--1169},
  url = {https://doi.org/10.1051/0004-6361:20011394}
}

@article{Kluge2025,
  author = {Kluge, M. and Hatch, N. A. and Montes, M. and others},
  title = {{Euclid: Early Release Observations. The intracluster light and intracluster globular clusters of the Perseus cluster}},
  journal = {Astronomy \& Astrophysics},
  year = {2025},
  volume = {697},
  eid = {A13},
  doi = {10.1051/0004-6361/202450772},
  pages = {A13},
  url = {https://doi.org/10.1051/0004-6361/202450772}
}

@article{Koda2015,
  author = {Koda, Jin and Yagi, Masafumi and Yamanoi, Hitomi and Komiyama, Yutaka},
  title = {{Approximately A Thousand Ultra Diffuse Galaxies in the Coma cluster}},
  year = {2015},
  eprint = {1506.01712},
  archivePrefix = {arXiv},
  journal = {The Astrophysical Journal Letters},
  volume = {807},
  pages = {L2},
  doi = {10.1088/2041-8205/807/1/L2},
  url = {https://doi.org/10.1088/2041-8205/807/1/L2}
}

@article{Laureijs1987,
  author = {{Laureijs}, R. J. and {Mattila}, K. and {Schnur}, G.},
  title = {{IRAS and optical observations of the high-latitude dust cloud Lynds 1642}},
  journal = {Astronomy \& Astrophysics},
  year = {1987},
  volume = {184},
  pages = {269--278}
}

@article{Lei2025,
  author = {Lei, Minjie and Clark, S. E. and Morel, Rudy and Allys, E. and Butsky, Iryna S. and Redshaw, Caleb and Fielding, Drummond B.},
  title = {{Neutral gas phase distribution from HI morphology: phase separation with scattering spectra and variational autoencoders}},
  year = {2025},
  eprint = {2505.20407},
  archivePrefix = {arXiv},
  journal = {The Astrophysical Journal},
  volume = {993},
  pages = {4},
  doi = {10.3847/1538-4357/ae044a},
  url = {https://doi.org/10.3847/1538-4357/ae044a}
}

@article{LeiClark2023,
  author = {Lei, Minjie and Clark, S. E.},
  title = {{Probing the cold neutral medium through HI emission morphology with the scattering transform}},
  year = {2023},
  eprint = {2212.06182},
  archivePrefix = {arXiv},
  doi = {10.3847/1538-4357/acc02a},
  journal = {The Astrophysical Journal},
  volume = {947},
  pages = {74},
  url = {https://doi.org/10.3847/1538-4357/acc02a}
}

@article{Li2022,
  author = {Li, Jiaxuan and Huang, Song and Leauthaud, Alexie and others},
  title = {{Reaching for the Edge I: Probing the Outskirts of Massive Galaxies with HSC, DECaLS, SDSS, and Dragonfly}},
  year = {2022},
  eprint = {2111.03557},
  archivePrefix = {arXiv},
  doi = {10.1093/mnras/stac2121},
  journal = {Monthly Notices of the Royal Astronomical Society},
  volume = {515},
  pages = {5335--5357},
  url = {https://doi.org/10.1093/mnras/stac2121}
}

@article{Liu2022,
  author = {Liu, Qing and Abraham, Roberto and Gilhuly, Colleen and others},
  title = {{A Method to Characterize the Wide-angle Point-Spread Function of Astronomical Images}},
  journal = {The Astrophysical Journal},
  year = {2022},
  volume = {925},
  eid = {219},
  doi = {10.3847/1538-4357/ac32c6},
  eprint = {2110.11598},
  archivePrefix = {arXiv},
  pages = {219},
  url = {https://doi.org/10.3847/1538-4357/ac32c6}
}

@article{Liu2023,
  author = {Liu, Qing and Abraham, Roberto G. and Martin, Peter G. and others},
  title = {{A Recipe for Unbiased Background Modeling in Deep Wide-field Astronomical Images}},
  journal = {The Astrophysical Journal},
  year = {2023},
  volume = {953},
  eid = {7},
  doi = {10.3847/1538-4357/acdee3},
  eprint = {2306.07922},
  archivePrefix = {arXiv},
  pages = {7},
  url = {https://doi.org/10.3847/1538-4357/acdee3}
}

@article{Liu2025a,
  author = {Liu, Qing and Abraham, Roberto and Martin, Peter G. and others},
  title = {{{Fuzzy Galaxies or Cirrus? Decomposition of Galactic Cirrus in Deep Wide-field Images}}},
  journal = {The Astrophysical Journal},
  year = {2025},
  volume = {979},
  number = {2},
  eid = {175},
  pages = {175},
  doi = {10.3847/1538-4357/ad9b25},
  archivePrefix = {arXiv},
  eprint = {2412.00933},
  primaryClass = {astro-ph.GA},
  adsurl = {https://ui.adsabs.harvard.edu/abs/2025ApJ...979..175L},
  url = {https://doi.org/10.3847/1538-4357/ad9b25}
}

@article{Liu2025b,
  author = {Liu, Qing and Martin, Peter and Abraham, Roberto and others},
  title = {{Quantitative morphology of galactic cirrus in deep optical imaging. Statistical structural analysis in a multiwavelength perspective}},
  journal = {Astronomy \& Astrophysics},
  year = {2025},
  volume = {704},
  eid = {A269},
  doi = {10.1051/0004-6361/202556295},
  eprint = {2507.05217},
  archivePrefix = {arXiv},
  pages = {A269},
  url = {https://doi.org/10.1051/0004-6361/202556295}
}

@article{Marchuk2021,
  author = {{Marchuk}, A. A. and {Smirnov}, A. A. and {Mosenkov}, A. V. and {Il'in}, V. B. and {Gontcharov}, G. A. and {Savchenko}, S. S. and {Rom{\'a}n}, J.},
  title = {{Fractal dimension of optical cirrus in {Stripe82}}},
  journal = {Monthly Notices of the Royal Astronomical Society},
  year = {2021},
  volume = {508},
  pages = {5825--5841},
  doi = {10.1093/mnras/stab2846},
  eprint = {2109.14034},
  archivePrefix = {arXiv},
  url = {https://doi.org/10.1093/mnras/stab2846}
}

@article{Marleau2025,
  author = {Marleau, F. R. and Cuillandre, J. -C. and Cantiello, M. and others},
  title = {{Euclid: Early Release Observations -- Dwarf galaxies in the Perseus galaxy cluster}},
  year = {2025},
  eprint = {2405.13502},
  archivePrefix = {arXiv},
  doi = {10.1051/0004-6361/202450799},
  journal = {Astronomy \& Astrophysics},
  volume = {697},
  pages = {A12},
  url = {https://doi.org/10.1051/0004-6361/202450799}
}

@article{Martin2010,
  author = {Martin, P. G. and Miville-Deschenes, M. -A. and Roy, A. and others},
  title = {{Direct Estimate of Cirrus Noise in Herschel Hi-GAL Images}},
  year = {2010},
  eprint = {1005.3076},
  archivePrefix = {arXiv},
  doi = {10.1051/0004-6361/201014684},
  journal = {Astronomy \& Astrophysics},
  volume = {518},
  pages = {L105},
  url = {https://doi.org/10.1051/0004-6361/201014684}
}

@article{MartínezDelgado2025,
  author = {Martinez-Delgado, David and Stein, Michael and Sakowska, Joanna D. and others},
  title = {{Stellar tidal streams around nearby spiral galaxies with deep imaging from amateur telescopes}},
  year = {2025},
  eprint = {2504.02071},
  archivePrefix = {arXiv},
  doi = {10.1051/0004-6361/202554980},
  journal = {Astronomy \& Astrophysics},
  volume = {701},
  pages = {A182},
  url = {https://doi.org/10.1051/0004-6361/202554980}
}

@article{McDonald2011,
  author = {McDonald, Michael and Courteau, Stephane and Tully, R. Brent and Roediger, Joel},
  title = {{A Survey of 286 Virgo Cluster Galaxies at Optical griz and Near-IR H-band: Surface Brightness Profiles and Bulge-Disk Decompositions}},
  year = {2011},
  eprint = {1105.0002},
  archivePrefix = {arXiv},
  doi = {10.1111/j.1365-2966.2011.18519.x},
  journal = {Monthly Notices of the Royal Astronomical Society},
  volume = {414},
  pages = {2055--2068},
  url = {https://doi.org/10.1111/j.1365-2966.2011.18519.x}
}

@article{Meisner2014,
  author = {Meisner, Aaron and Finkbeiner, Douglas},
  title = {{A Full-sky, High-resolution Atlas of Galactic 12 micron Dust Emission with WISE}},
  year = {2014},
  eprint = {1312.0947},
  archivePrefix = {arXiv},
  doi = {10.1088/0004-637X/781/1/5},
  journal = {The Astrophysical Journal},
  volume = {781},
  pages = {5},
  url = {https://doi.org/10.1088/0004-637X/781/1/5}
}

@article{Merritt2016,
  author = {Merritt, Allison and van Dokkum, Pieter and Abraham, Roberto and Zhang, Jielai},
  title = {{The Dragonfly Nearby Galaxies Survey. I. Substantial variation in the diffuse stellar halos around spiral galaxies}},
  year = {2016},
  eprint = {1606.08847},
  archivePrefix = {arXiv},
  doi = {10.3847/0004-637X/830/2/62},
  journal = {The Astrophysical Journal},
  volume = {830},
  pages = {62},
  url = {https://doi.org/10.3847/0004-637X/830/2/62}
}

@article{Mihos2005,
  author = {Mihos, Chris and Harding, Paul and Feldmeier, John and Morrison, Heather},
  title = {{Diffuse Light in the Virgo Cluster}},
  year = {2005},
  eprint = {astro-ph/0508217},
  archivePrefix = {arXiv},
  doi = {10.1086/497030},
  journal = {The Astrophysical Journal Letters},
  volume = {631},
  pages = {L41--L44},
  url = {https://doi.org/10.1086/497030}
}

@article{Mihos2017,
  author = {Mihos, J. Christopher and Harding, Paul and Feldmeier, John J. and Rudick, Craig and Janowiecki, Steven and Morrison, Heather and Slater, Colin and Watkins, Aaron},
  title = {{The Burrell Schmidt Deep Virgo Survey: Tidal Debris, Galaxy Halos, and Diffuse Intracluster Light in the Virgo Cluster}},
  journal = {The Astrophysical Journal},
  year = {2017},
  volume = {834},
  eid = {16},
  doi = {10.3847/1538-4357/834/1/16},
  eprint = {1611.04435},
  archivePrefix = {arXiv},
  pages = {16},
  url = {https://doi.org/10.3847/1538-4357/834/1/16}
}

@misc{Mihos2019,
  author = {Mihos, J. Christopher},
  title = {{Deep Imaging of Diffuse Light Around Galaxies and Clusters: Progress and Challenges}},
  year = {2019},
  eprint = {1909.09456},
  archivePrefix = {arXiv},
  primaryClass = {astro-ph.GA},
  howpublished = {arXiv e-prints},
  adsurl = {https://ui.adsabs.harvard.edu/abs/2019arXiv190909456M/abstract},
  url = {https://arxiv.org/abs/1909.09456}
}

@ARTICLE{MivilleDeschenes2003,
       author = {{Miville-Desch{\^e}nes}, M.-A. and {Levrier}, F. and {Falgarone}, E.},
        title = "{On the Use of Fractional Brownian Motion Simulations to Determine the Three-dimensional Statistical Properties of Interstellar Gas}",
      journal = {\apj},
         year = 2003,
        month = aug,
       volume = {593},
       number = {2},
        pages = {831-847},
          doi = {10.1086/376603},
archivePrefix = {arXiv},
       eprint = {astro-ph/0304539},
 primaryClass = {astro-ph},
       adsurl = {https://ui.adsabs.harvard.edu/abs/2003ApJ...593..831M}
}

@article{MivilleDeschenes2007,
  author = {{Miville-Desch{\^e}nes}, M.-A. and {Lagache}, G. and {Boulanger}, F. and {Puget}, J.-L.},
  title = {{Statistical properties of dust far-infrared emission}},
  journal = {Astronomy \& Astrophysics},
  year = {2007},
  volume = {469},
  pages = {595--605},
  doi = {10.1051/0004-6361:20066962},
  eprint = {0704.2175},
  archivePrefix = {arXiv},
  url = {https://doi.org/10.1051/0004-6361:20066962}
}

@article{MivilleDeschenes2016,
  author = {{Miville-Desch\^enes}, M.-A. and {Duc}, P.-A. and {Marleau}, F. and {Cuillandre}, J.-C. and {Didelon}, P. and {Gwyn}, S. and {Karabal}, E.},
  title = {{Probing interstellar turbulence in cirrus with deep optical imaging: no sign of energy dissipation at 0.01 pc scale}},
  journal = {Astronomy \& Astrophysics},
  year = {2016},
  volume = {593},
  eid = {A4},
  doi = {10.1051/0004-6361/201628503},
  eprint = {1605.08360},
  archivePrefix = {arXiv},
  pages = {A4},
  url = {https://doi.org/10.1051/0004-6361/201628503}
}

@article{MontenegroTaborda2023,
  author = {Montenegro-Taborda, Daniel and Rodriguez-Gomez, Vicente and Pillepich, Annalisa and Avila-Reese, Vladimir and Sales, Laura V. and Rodr{\'{i}}guez-Puebla, Aldo and Hernquist, Lars},
  title = {{The growth of brightest cluster galaxies in the TNG300 simulation:dissecting the contributions from mergers and in situ star formation}},
  year = {2023},
  eprint = {2302.10943},
  archivePrefix = {arXiv},
  doi = {10.1093/mnras/stad586},
  journal = {Monthly Notices of the Royal Astronomical Society},
  volume = {521},
  pages = {800},
  url = {https://doi.org/10.1093/mnras/stad586}
}

@article{Montes2018,
  author = {Montes, Mireia and Trujillo, Ignacio},
  title = {{Intracluster light at the Frontier -- II. The Frontier Fields Clusters}},
  doi = {10.1093/mnras/stx2847},
  year = {2018},
  journal = {Monthly Notices of the Royal Astronomical Society},
  volume = {474},
  pages = {917--932},
  url = {https://doi.org/10.1093/mnras/stx2847}
}

@article{Montes2022,
  author = {Montes, Mireia},
  title = {{The faint light in groups and clusters of galaxies}},
  doi = {10.1038/s41550-022-01616-z},
  year = {2022},
  journal = {Nature Astronomy},
  volume = {6},
  pages = {308--316},
  url = {https://doi.org/10.1038/s41550-022-01616-z}
}

@article{Mowla2017,
  author = {Mowla, Lamiya and van Dokkum, Pieter and Merritt, Allison and Abraham, Roberto and Yagi, Masafumi and Koda, Jin},
  title = {{Evidence of Absence of Tidal Features in the Outskirts of Ultra Diffuse Galaxies in the Coma Cluster}},
  year = {2017},
  eprint = {1710.03762},
  archivePrefix = {arXiv},
  doi = {10.3847/1538-4357/aa961b},
  journal = {The Astrophysical Journal},
  volume = {851},
  pages = {27},
  url = {https://doi.org/10.3847/1538-4357/aa961b}
}

@article{Pillepich2014,
  author = {Pillepich, Annalisa and Vogelsberger, Mark and Deason, Alis and others},
  title = {{Halo Mass and Assembly History Exposed in the Faint Outskirts: the Stellar and Dark Matter Haloes of Illustris Galaxies}},
  year = {2014},
  eprint = {1406.1174},
  archivePrefix = {arXiv},
  doi = {10.1093/mnras/stu1408},
  journal = {Monthly Notices of the Royal Astronomical Society},
  volume = {444},
  pages = {237--249},
  url = {https://doi.org/10.1093/mnras/stu1408}
}

@article{Pillepich2018,
  author = {Pillepich, Annalisa and Nelson, Dylan and Hernquist, Lars and others},
  title = {{First results from the IllustrisTNG simulations: the stellar mass content of groups and clusters of galaxies}},
  year = {2018},
  eprint = {1707.03406},
  archivePrefix = {arXiv},
  doi = {10.1093/mnras/stx3112},
  journal = {Monthly Notices of the Royal Astronomical Society},
  volume = {475},
  pages = {648--675},
  url = {https://doi.org/10.1093/mnras/stx3112}
}

@article{PlanckCollaborationXXXII2016,
  author = {{Planck Collaboration}},
  title = {{Planck intermediate results. XXXII. The relative orientation between the magnetic field and structures traced by interstellar dust}},
  year = {2016},
  eprint = {1409.6728},
  archivePrefix = {arXiv},
  doi = {10.1051/0004-6361/201425044},
  journal = {Astronomy \& Astrophysics},
  volume = {586},
  pages = {A135},
  url = {https://doi.org/10.1051/0004-6361/201425044}
}

@article{Poliakov2026,
  author = {{Poliakov}, D. M. and {Smirnov}, A. A. and {Savchenko}, S. S. and {Marchuk}, A. A. and {Mosenkov}, A. V. and {Il'in}, V. B. and {Gontcharov}, G. A. and {Turichina}, D. G. and {Panasyuk}, A. D.},
  title = {{Analysis of Galactic cirrus filaments in {HSC-SSP} high-resolution deep images using artificial neural networks}},
  journal = {Astronomy and Computing},
  year = {2026},
  volume = {55},
  eid = {101075},
  doi = {10.1016/j.ascom.2026.101075},
  eprint = {2602.09779},
  archivePrefix = {arXiv},
  pages = {101075},
  url = {https://doi.org/10.1016/j.ascom.2026.101075}
}

@article{Postman2024,
  author = {Postman, Marc and Lauer, Tod R. and Parker, Joel W. and others},
  title = {{New Synoptic Observations of the Cosmic Optical Background with {New Horizons}}},
  journal = {The Astrophysical Journal},
  year = {2024},
  volume = {972},
  eid = {95},
  doi = {10.3847/1538-4357/ad5ffc},
  eprint = {2407.06273},
  archivePrefix = {arXiv},
  pages = {95},
  url = {https://doi.org/10.3847/1538-4357/ad5ffc}
}

@article{RodriguezGomez2016,
  author = {Rodriguez-Gomez, Vicente and Pillepich, Annalisa and Sales, Laura V. and others},
  title = {{The stellar mass assembly of galaxies in the Illustris simulation: growth by mergers and the spatial distribution of accreted stars}},
  year = {2016},
  eprint = {1511.08804},
  archivePrefix = {arXiv},
  doi = {10.1093/mnras/stw456},
  journal = {Monthly Notices of the Royal Astronomical Society},
  volume = {458},
  pages = {2371--2390},
  url = {https://doi.org/10.1093/mnras/stw456}
}

@article{Roman2020,
  author = {{Rom{\'a}n}, J. and {Trujillo}, I. and {Montes}, M.},
  title = {{Galactic cirri in deep optical imaging}},
  journal = {Astronomy \& Astrophysics},
  year = {2020},
  volume = {644},
  eid = {A42},
  doi = {10.1051/0004-6361/201936111},
  eprint = {1907.00978},
  archivePrefix = {arXiv},
  pages = {A42},
  url = {https://doi.org/10.1051/0004-6361/201936111}
}

@article{Sandage1976,
  author = {Sandage, Allan},
  title = {{High-Latitude Reflection Nebulosities Illuminated by the Galactic Plane}},
  doi = {10.1086/111975},
  year = {1976},
  journal = {The Astronomical Journal},
  volume = {81},
  pages = {954--957},
  url = {https://doi.org/10.1086/111975}
}

@article{Sandin2014,
  author = {{Sandin}, C.},
  title = {{The influence of diffuse scattered light. I. The {PSF} and its role in observations of the edge-on galaxy {NGC 5907}}},
  journal = {Astronomy \& Astrophysics},
  year = {2014},
  volume = {567},
  eid = {A97},
  doi = {10.1051/0004-6361/201423429},
  eprint = {1406.5508},
  archivePrefix = {arXiv},
  pages = {A97},
  url = {https://doi.org/10.1051/0004-6361/201423429}
}

@article{Saury2014,
  author = {Saury, E. and Miville-Desch{\^{e}}nes, M. -A. and Hennebelle, P. and Audit, E. and Schmidt, W.},
  title = {{The structure of the thermally bistable and turbulent atomic gas in the local interstellar medium}},
  year = {2014},
  eprint = {1301.3446},
  archivePrefix = {arXiv},
  doi = {10.1051/0004-6361/201321113},
  journal = {Astronomy \& Astrophysics},
  volume = {567},
  pages = {A16},
  url = {https://doi.org/10.1051/0004-6361/201321113}
}

@article{Saydjari2022,
  author = {Saydjari, Andrew K. and Finkbeiner, Douglas P.},
  title = {{Photometry on Structured Backgrounds: Local Pixelwise Infilling by Regression}},
  year = {2022},
  eprint = {2201.07246},
  archivePrefix = {arXiv},
  doi = {10.3847/1538-4357/ac6875},
  journal = {The Astrophysical Journal},
  volume = {933},
  pages = {155},
  url = {https://doi.org/10.3847/1538-4357/ac6875}
}

@article{Simon2019,
  author = {Simon, Joshua D.},
  title = {{The Faintest Dwarf Galaxies}},
  year = {2019},
  eprint = {1901.05465},
  archivePrefix = {arXiv},
  doi = {10.1146/annurev-astro-091918-104453},
  journal = {Annual Review of Astronomy and Astrophysics},
  volume = {57},
  pages = {375--415},
  url = {https://doi.org/10.1146/annurev-astro-091918-104453}
}

@article{Sinha2026,
  author = {{Sinha}, S. and {Ghosh}, T. and {Allys}, E. and {Boulanger}, F. and {Delouis}, J.-M.},
  title = {{Untangling dust emission and cosmic infrared background anisotropies with the scattering transform statistics}},
  journal = {Astronomy \& Astrophysics},
  year = {2026},
  volume = {711},
  eid = {A287},
  doi = {10.1051/0004-6361/202659670},
  eprint = {2603.06110},
  archivePrefix = {arXiv},
  pages = {A287},
  url = {https://doi.org/10.1051/0004-6361/202659670}
}

@article{Slater2009,
  author = {{Slater}, Colin T. and {Harding}, Paul and {Mihos}, J. Christopher},
  title = {{{Removing Internal Reflections from Deep Imaging Data Sets}}},
  journal = {Publications of the Astronomical Society of the Pacific},
  year = {2009},
  volume = {121},
  number = {885},
  pages = {1267},
  doi = {10.1086/648457},
  archivePrefix = {arXiv},
  eprint = {0909.3320},
  primaryClass = {astro-ph.IM},
  adsurl = {https://ui.adsabs.harvard.edu/abs/2009PASP..121.1267S},
  url = {https://doi.org/10.1086/648457}
}

@article{Smirnov2023,
  author = {{Smirnov}, A. A. and {Savchenko}, S. S. and {Poliakov}, D. M. and {Marchuk}, A. A. and {Mosenkov}, A. V. and {Il'in}, V. B. and {Gontcharov}, G. A. and {Rom{\'a}n}, J. and {Seguine}, J.},
  title = {{Prospects for future studies using deep imaging: analysis of individual Galactic cirrus filaments}},
  journal = {Monthly Notices of the Royal Astronomical Society},
  year = {2023},
  volume = {519},
  pages = {4735--4752},
  doi = {10.1093/mnras/stac3765},
  eprint = {2301.12410},
  archivePrefix = {arXiv},
  url = {https://doi.org/10.1093/mnras/stac3765}
}

@article{Sola2025,
  author = {Sola, Elisabeth and Duc, Pierre-Alain and Urbano, Mathias and others},
  title = {{Low Surface Brightness structures from annotated deep CFHT images: effects of the host galaxy's properties and environment}},
  year = {2025},
  eprint = {2503.18480},
  archivePrefix = {arXiv},
  doi = {10.1093/mnras/staf1139},
  journal = {Monthly Notices of the Royal Astronomical Society},
  volume = {541},
  pages = {3015--3042},
  url = {https://doi.org/10.1093/mnras/staf1139}
}

@article{Trujillo2016,
  author = {Trujillo, Ignacio and Fliri, Juergen},
  title = {{Beyond 31 mag/arcsec^2: the low surface brightness frontier with the largest optical telescopes}},
  year = {2016},
  eprint = {1510.04696},
  archivePrefix = {arXiv},
  doi = {10.3847/0004-637X/823/2/123},
  journal = {The Astrophysical Journal},
  volume = {823},
  pages = {123},
  url = {https://doi.org/10.3847/0004-637X/823/2/123}
}

@article{vanDokkum2015,
  author = {van Dokkum, Pieter and Abraham, Roberto and Merritt, Allison and Zhang, Jielai and Geha, Marla and Conroy, Charlie},
  title = {{Forty-Seven Milky Way-Sized, Extremely Diffuse Galaxies in the Coma Cluster}},
  year = {2015},
  eprint = {1410.8141},
  archivePrefix = {arXiv},
  doi = {10.1088/2041-8205/798/2/L45},
  journal = {The Astrophysical Journal Letters},
  volume = {798},
  pages = {L45},
  url = {https://doi.org/10.1088/2041-8205/798/2/L45}
}

@article{Venhola2017,
  author = {Venhola, A. and Peletier, R. and Laurikainen, E. and others},
  title = {{The Fornax Deep Survey (FDS) with the VST: III. Low Surface Brightness (LSB) dwarfs and Ultra Diffuse Galaxies (UDGs) in the center of the Fornax cluster}},
  year = {2017},
  eprint = {1710.04616},
  archivePrefix = {arXiv},
  doi = {10.1051/0004-6361/201730696},
  journal = {Astronomy \& Astrophysics},
  volume = {608},
  pages = {A142},
  url = {https://doi.org/10.1051/0004-6361/201730696}
}

@article{Viero2013,
  author = {Viero, M. P. and Wang, L. and Zemcov, M. and others},
  title = {{HerMES: Cosmic Infrared Background Anisotropies and the Clustering of Dusty Star-Forming Galaxies}},
  year = {2013},
  eprint = {1208.5049},
  archivePrefix = {arXiv},
  doi = {10.1088/0004-637X/772/1/77},
  journal = {The Astrophysical Journal},
  volume = {772},
  pages = {77},
  url = {https://doi.org/10.1088/0004-637X/772/1/77}
}

@article{Watkins2024,
  author = {{Watkins}, A. E. and {Kaviraj}, S. and {Collins}, C. C. and {Knapen}, J. H. and {Kelvin}, L. S. and {Duc}, P.-A. and {Rom{\'a}n}, J. and {Mihos}, J. C.},
  title = {{Strategies for optimal sky subtraction in the low surface brightness regime}},
  journal = {Monthly Notices of the Royal Astronomical Society},
  year = {2024},
  volume = {528},
  pages = {4289--4306},
  doi = {10.1093/mnras/stae236},
  eprint = {2401.12297},
  archivePrefix = {arXiv},
  url = {https://doi.org/10.1093/mnras/stae236}
}

@article{Witt2008,
  author = {{Witt}, A. N. and {Mandel}, S. and {Sell}, P. H. and {Dixon}, T. and {Vijh}, U. P.},
  title = {{Extended Red Emission in High-Galactic-Latitude Interstellar Clouds}},
  journal = {The Astrophysical Journal},
  year = {2008},
  volume = {679},
  pages = {497--511},
  doi = {10.1086/587131},
  eprint = {0802.0674},
  archivePrefix = {arXiv},
  url = {https://doi.org/10.1086/587131}
}

@article{Zaritsky2019,
  author = {Zaritsky, Dennis and Donnerstein, Richard and Dey, Arjun and Kadowaki, Jennifer and Zhang, Huanian and Karunakaran, Ananthan and Mart{\'{i}}nez-Delgado, David and Rahman, Mubdi and Spekkens, Kristine},
  title = {{Systematically Measuring Ultra Diffuse Galaxies (SMUDGes). I. Survey Description and First Results in the Coma Galaxy Cluster and Environs}},
  year = {2019},
  eprint = {1811.04098},
  archivePrefix = {arXiv},
  doi = {10.3847/1538-4365/aaefe9},
  journal = {The Astrophysical Journal Supplement Series},
  volume = {240},
  pages = {1},
  url = {https://doi.org/10.3847/1538-4365/aaefe9}
}

@article{Zaritsky2024,
  author = {Zaritsky, Dennis and Golini, Giulia and Donnerstein, Richard and others},
  title = {{LIGHTS. Survey Overview and a Search for Low Surface Brightness Satellite Galaxies}},
  year = {2024},
  eprint = {2406.01912},
  archivePrefix = {arXiv},
  journal = {The Astronomical Journal},
  volume = {168},
  pages = {69},
  doi = {10.3847/1538-3881/ad543f},
  url = {https://doi.org/10.3847/1538-3881/ad543f}
}

@ARTICLE{Zaroubi1995,
       author = {{Zaroubi}, S. and {Hoffman}, Y. and {Fisher}, K.~B. and {Lahav}, O.},
        title = "{Wiener Reconstruction of the Large-Scale Structure}",
      journal = {\apj},
         year = 1995,
        month = aug,
       volume = {449},
        pages = {446},
          doi = {10.1086/176070},
archivePrefix = {arXiv},
       eprint = {astro-ph/9410080},
 primaryClass = {astro-ph},
       adsurl = {https://ui.adsabs.harvard.edu/abs/1995ApJ...449..446Z}
}

@article{Zemcov2017,
  author = {Zemcov, Michael and Immel, Poppy and Nguyen, Chi and Cooray, Asantha and Lisse, Carey M. and Poppe, Andrew R.},
  title = {{Measurement of the cosmic optical background using the Long Range Reconnaissance Imager on New Horizons}},
  journal = {Nature Communications},
  year = {2017},
  volume = {8},
  eid = {15003},
  doi = {10.1038/ncomms15003},
  eprint = {1704.02989},
  archivePrefix = {arXiv},
  pages = {15003},
  url = {https://doi.org/10.1038/ncomms15003}
}

@article{Zhang2023,
  author = {{Zhang}, J. and {Martin}, P. G. and {Cloutier}, R. and {Price-Jones}, N. and {Abraham}, R. and {van Dokkum}, P. and {Merritt}, A.},
  title = {{Joint Modelling of Dust Scattering and Thermal Emission: The Spider Complex}},
  journal = {The Astrophysical Journal},
  year = {2023},
  volume = {948},
  eid = {4},
  doi = {10.3847/1538-4357/acc177},
  eprint = {2305.00756},
  archivePrefix = {arXiv},
  pages = {4},
  url = {https://doi.org/10.3847/1538-4357/acc177}
}

@article{Zhao2024,
  author = {{Zhao}, Y. and {Zhang}, W. and {Ma}, L. and {Wen}, S. and {Wu}, H.},
  title = {{Galactic Cirri at High Galactic Latitudes. I. Investigating Scatter in Slopes between Optical and Far-infrared Intensities}},
  journal = {The Astronomical Journal},
  year = {2024},
  volume = {168},
  eid = {88},
  doi = {10.3847/1538-3881/ad58d5},
  eprint = {2406.03031},
  archivePrefix = {arXiv},
  pages = {88},
  url = {https://doi.org/10.3847/1538-3881/ad58d5}
}

@article{Yoo2024Spatial5,
  title = {{Spatial Distribution of Intracluster Light versus Dark Matter in Horizon Run 5}},
  year = {2024},
  author = {Yoo, Jaewon and Park, Changbom and Sabiu, Cristiano G. and Singh, Ankit and Ko, Jongwan and Lee, Jaehyun and Pichon, Christophe and Jee, M. James and Gibson, Brad K. and Snaith, Owain and Kim, Juhan and Shin, Jihye and Kim, Yonghwi and Kim, Hyowon},
  url = {http://arxiv.org/abs/2402.17958},
  arxivid = {2402.17958},
}

\appendix

\section{Injection tests with synthetic cirrus}
\label{app:mock_recovery}

\subsection{Mock cirrus fields and halo injections}

\begin{figure*}[!htbp]
\centering
  \resizebox{\hsize}{!}{\includegraphics{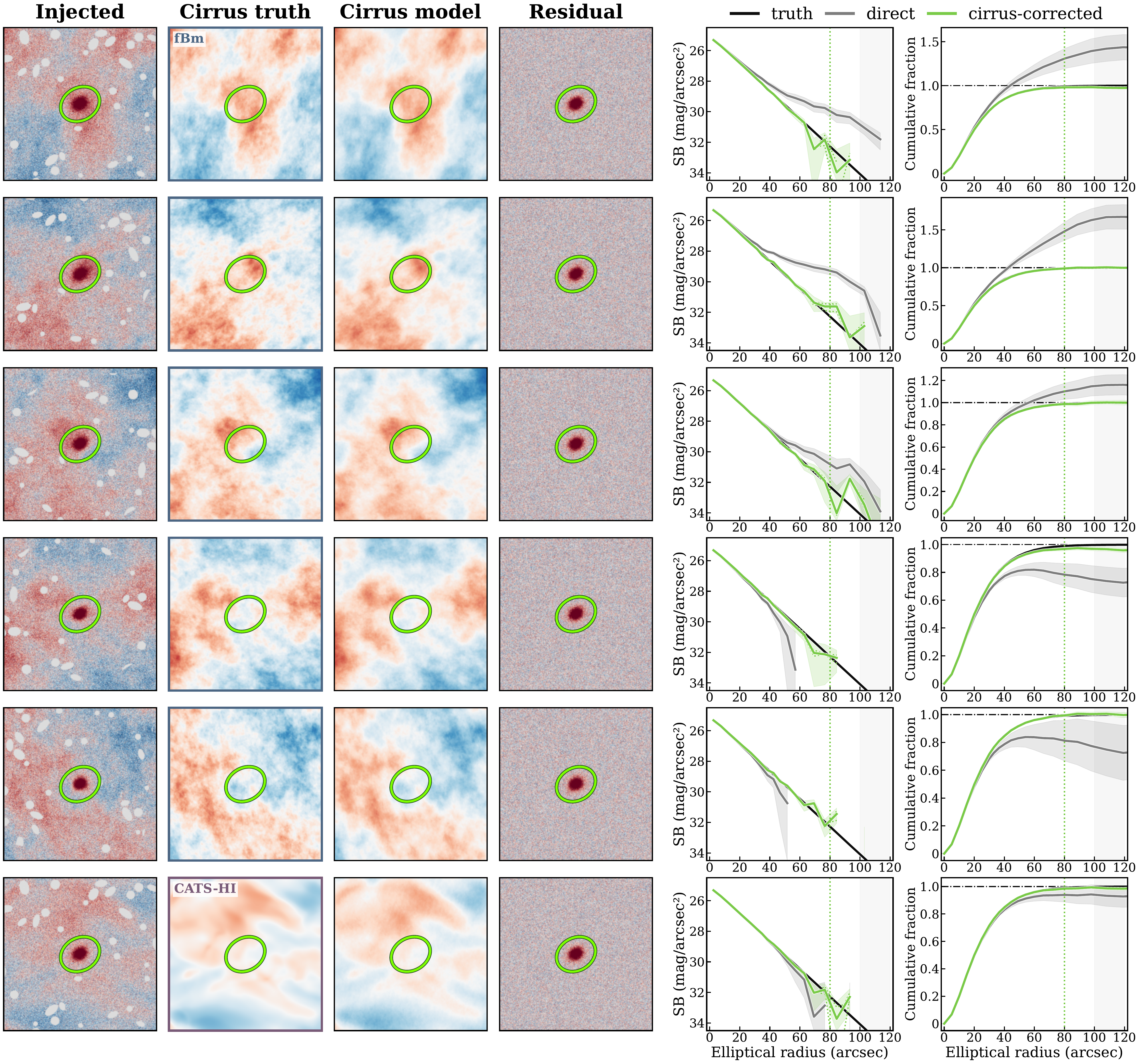}}
  \caption{Halo recovery tests with synthetic Galactic cirrus. Each row shows the central cutout of a mock cirrus field (1-5th row: anisotropic fBm fields; 6th row: an MHD-based mock field) containing an injected exponential halo ($n=1$) with $R_{\rm e}=20\arcsec$. From left to right column: injected image, cirrus truth, mean reconstructed cirrus from three recovery draws, mean cirrus-subtracted residual, radial surface brightness profiles, and cumulative flux growth curves. The masked pixels in gray and the pixels within the $4\,R_{\rm e}$ ellipse shown in the image panels are excluded from the modeling. In the profile panels, black, gray, and green curves indicate truths, direct measurements, and cirrus-corrected results, respectively. Dotted green curves show individual realizations. The shaded gray band indicate the sky annulus. The cumulative flux is normalized by the total injected flux.}
\label{fig:mock_recovery}
\end{figure*}

We tested the cirrus modeling approach using mock cirrus fields, where the foreground and injection are known. We consider two types of synthetic cirrus morphology:

First, we generated anisotropic fBm realizations with a 2D power spectrum following:
\begin{equation}
    P(\boldsymbol{k})\propto \boldsymbol{k}^{\beta},
\end{equation}
where $\beta$ is the power index. 
This power-law behavior matches that observed for diffuse high-latitude Galactic dust \citep[e.g.,][]{MivilleDeschenes2007,Martin2010}, with $\beta\simeq-3$ for optical cirrus \citep{MivilleDeschenes2016,Liu2025b}.
The mock fields have different random phases, orientations, and degrees of anisotropy. 

Second, we constructed mock fields from a synthetic {\HI} column-density map provided by the Catalogue for Astrophysical Turbulence Simulations (CATS; \citealt{Burkhart2020}). The map was derived from isothermal magnetohydrodynamic (MHD) simulations of turbulent gas.
We adopted a simulation with sonic and Alfv\'enic Mach numbers of $M_{\rm s}=4.5$ and $M_{\rm A}=0.7$, respectively.
Unlike the Gaussian fBm fields, this map contains coherent structures strongly driven by the turbulence, providing an additional test of foreground pattern.
For the MHD-based mocks, we assumed that the optical cirrus intensity is proportional to the {\HI} column density. The map was resampled onto the VIS image grid
and PSF convolved.\footnote{At the adopted angular scaling, the original CATS-H\,I map has a sampling of $\sim6\arcsec$ per pixel. Resampling to the $0.9\arcsec$ VIS grid does not introduce additional small-scale structure.}
Both types of cirrus fields are normalized to a mean surface brightness of $\mu_{\rm cir}=26.5$ mag arcsec$^{-2}$.

The mock image has the same $10\arcmin \times 10\arcmin$ field-of-view and a pixel resolution of $0 \farcs 9$ as in the main experiments.
In each field, we injected an exponential halo with $R_{\rm e}=20\arcsec$, $\mu_0=25$ mag/arcsec$^2$, and axis ratio $q=0.8$. 
The image was then convolved with the \textit{Euclid} VIS PSF.
We then added instrumental noise matching the \textit{Euclid} ERO VIS observations using the surface brightness depths reported in \cite{Cuillandre2025a}. For photometry, we used a sky annulus of $100\arcsec$--$140\arcsec$. We randomly placed source masks in the images as elliptical apertures. 

For these tests, the FIR prior was generated directly from the known cirrus field, convolved to a beam FWHM of $18\arcsec$, which is twice as sharp as the beam used in the main experiments. This is a more idealized guide than those used for the real data: no CIB anisotropies persist and the optical-FIR ratio is strictly fixed. Therefore, we reduced the information available by excluding all pixels in the $80\arcsec$ central ellipse aperture from the reconstruction. 
We applied the same inference framework following the main experiments. Three independent reconstructions were optimized for each field. 

\subsection{Halo light recovery in mock cirrus fields}

Figure~\ref{fig:mock_recovery} illustrates the results with five example fBm realizations and one MHD-based mock. The first three rows illustrate cases where cirrus contamination produces a positive excess in the halo outskirts, leading to overestimated aperture fluxes. The bottom three rows instead show where the measured profiles decline too steeply because of bias in the local sky estimate.
The reconstructed cirrus closely follows the cirrus truth. The radial profiles and cumulative flux measurements show significant improvements across fields with positive and negative bias. 
Over ten mock fBm fields, the mean absolute
flux bias within $4\,R_{\rm e}$ ($80\arcsec$) decreases from $21\%$ in direct measurements to $1\%$ after cirrus correction, with improvements in all fields; the RMS of the cumulative flux also decreases from $24\%$ to $1\%$. 
The remaining differences are marginally visible in the faint outskirts fainter than about 31 mag/arcsec$^2$.

These tests demonstrate the potential of statistical cirrus modeling when the broad foreground is well informed. The mock setups benefited from the exact optical--FIR correspondence, known instrumental noise, and homogeneous multiscale statistics matching the foreground, and therefore produced optimistic results.
Real observations introduce additional uncertainties from FIR calibration, CIB contamination, intrinsic scatter in the optical--FIR relation, more complex noise patterns, and more complex cirrus morphology. The mock results therefore yield a controlled reference for the reconstruction, while the experiments using real data in Sec.~\ref{sec:halo} provide a more realistic assessment of its performance.

\section{Construction of the initial cirrus realizations}
\label{app:init_constrution}

\subsection{Conditional broad field initializations}
\label{app:init_broad}

\begin{figure}[!htbp]
\centering
  \resizebox{\hsize}{!}{\includegraphics{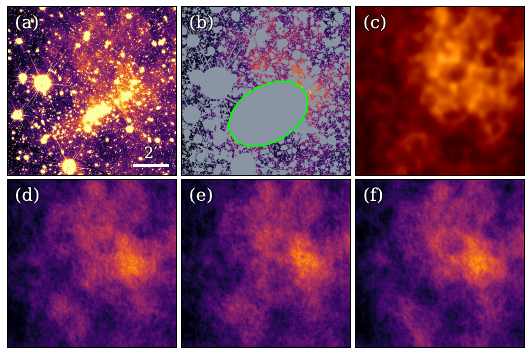}}
  \caption{Inputs and outputs of conditional stochastic initializations for cirrus synthesis in the central Abell 2390 field. (a) \textit{Euclid} VIS image. (b) VIS input for conditioning, with sources and the central elliptical region masked. (c) Herschel backbone, after sources cleaning and low pass filtering. (d–f) three example independent initialization realizations, $I_{{\rm cir},0}^{i}$ ($i=0,1,2$), out of the ten used for experiments in Fig.~\ref{fig:A2390_bcg_icl}.}
\label{fig:A2390_central_init}
\end{figure}

We first generate broad field initializations. The initialization uses the residual $\Delta_{\rm b}$ between the observed optical image and the FIR backbone, measured on unmasked pixels outside the target ellipse.
We describe this pattern with a stationary Matérn-$3/2$ Gaussian field, whose 2D power spectrum is:
\begin{equation}
P_{\rm b}(\boldsymbol{k}) \propto
\left(
\frac{3}{\ell_{\rm b}^{2}}+|\boldsymbol{k}|^{2}
\right)^{-5/2},
\label{eq:app_broad_spectrum}
\end{equation}
where $\ell_{\rm b}=60$ pixels is the correlation length. This covariance describes spatially correlated departures from the FIR backbone.
The conditioning includes a nuisance covariance $\mathbf{K}_{\eta}$, combining the measured instrumental noise covariance with a Gaussian component of correlation scale $\ell_{\rm s}=12$ pixels. The latter accounts for local optical fluctuations not described by $\mathbf{K}_{\rm b}$, whose scale is chosen to be consistent with the Gaussian scale separation in Eq.~(\ref{eq:cirrus_initial}).
The covariance is estimated from the exterior optical residuals after mask-normalized smoothing and accounting for the expected instrumental noise.

Let $O$ denote the source-masked observed pixels outside the target ellipse, and $X$ the full reconstruction domain. 
The conditional mean of the broad optical--FIR discrepancy is given by a Wiener-filter estimate \citep{Zaroubi1995}:
\begin{equation}
\overline{\Delta}_{\rm b}^{X} =
\mathbf{K}_{\rm b}^{XO}
\mathbf{C}_{OO}^{-1}
\left(I_{\rm obs}-I_{\rm bb}\right)^{O},
\label{eq:app_broad_condition}
\end{equation}
where $\mathbf{K}_{\rm b}^{XO}$ describes the covariance between the reconstruction and conditioning pixels, $\mathbf{K}_{\rm b}^{OO}$ describes the covariance among the conditioning pixels, and $\mathbf{C}_{OO} = \mathbf{K}_{\rm b}^{OO}+\mathbf{K}_{\eta}^{OO}$.

We generate conditional realizations following the constrained-realization algorithm of \citet{HoffmanRibak1991}.\footnote{Constrained realizations originate in cosmology, where they are used to generate density fields consistent with sparse, noisy, and incomplete survey data. We use the Gaussian description only for the broad component; the non-Gaussian, filamentary structure of the cirrus is constrained by the multiscale statistics
(Sec.~\ref{sec:method_stats}).} 
For each realization $m$, we draw independent Gaussian fields 
$u_{\rm b}^{m}\sim\mathcal{N}(0,\mathbf{K}_{\rm b})$ and
$u_{\eta}^{m}\sim\mathcal{N}(0,\mathbf{K}_{\eta})$.
The conditional broad discrepancy for realization $m$ is therefore:
\begin{equation}
\begin{split}
\Delta_{\rm b}^{m,X}
= {} & \overline{\Delta}_{\rm b}^{X}
+u_{\rm b}^{m,X} - \mathbf{K}_{\rm b}^{XO}\mathbf{C}_{OO}^{-1}
\left(u_{\rm b}^{m}+u_{\eta}^{m}\right)^O,
\end{split}
\label{eq:app_conditional_draw}
\end{equation}
where the last term is the conditional correction. Under the adopted Gaussian model, each realization is a draw from the distribution of $\Delta_{\rm b}$ conditioned on the exterior pixels, scattering around the conditional mean of Eq.~(\ref{eq:app_conditional_draw}) with the conditional covariance. 

Figure~\ref{fig:A2390_central_init} illustrates the constructed conditional initializations for the central Abell~2390 field. Panel (a) is the original VIS image. Panel (b) is the source and target masked input. The FIR backbone shown in panel (c) serves as a common prior. Three example independent conditional initializations are illustrated in the bottom panels. These realizations show differences in their small-scale structures allowed by the FIR and broad-field prior, which provide different starting points for the optimization.

\subsection{Locally constrained stochastic texture}
\label{app:local_texture}

Next, we generate the small-scale texture initialization $T^m$ by estimating the cirrus fluctuations in different directions, at different spatial scales, and at different positions in the image.

We started from independent random-phase Gaussian fields with a power-law spectrum $P(k)\propto k^{\beta}$ sampled on the image grid. ${\beta}$ is evaluated by  comparing the measured power spectrum with power-law templates after accounting for instrumental noise and mask support, with $\beta$ spanning $-1.5$ to $-4.5$ at a step of 0.5. For Abell~2390 VIS data, we obtained ${\beta}=-2.5$. These independent fields are isotropic, i.e., local structure has no preferred direction.
To reproduce the local anisotropy, for each generated field we divide its Fourier modes into five spatial-frequency bands and adjust their anisotropy using the observed optical pixels. The bands correspond to the five wavelet scales used in the WST, with an overlapping window for smooth transition between scales.
At each scale $j$, we measure the position-dependent, noise-corrected angular wavelet energy $E_{j,\theta}(\boldsymbol{x})$ in eight orientations corresponding to that used in the WST.
The preferred Fourier direction is estimated from the moment:
\begin{equation}
\vartheta_j(\boldsymbol{x})
=\frac{1}{2}\arg\left[
\sum_{\theta}E_{j,\theta}(\boldsymbol{x})
e^{2i\varphi_\theta}
\right],
\label{eq:app_texture_orientation}
\end{equation}
where $\varphi_\theta$ is the Fourier orientation of the wavelet filter.
The corresponding filament axis in real space is perpendicular to this direction.\footnote{This is because of a cirrus filament has stronger Fourier response across its width than along its length. Therefore, the measured Fourier orientation is rotated by $90\degr$ to obtain a filament orientation estimate.}
Where the angular-energy contrast across orientations is strong, $\vartheta_j(\boldsymbol{x})$ determines the local anisotropy. Where the directional energy contrast is weak or is limited by the valid pixel support, the structure remains isotropic.

To account for the dependence of cirrus fluctuations on local
brightness, we modulate the texture amplitude using observed cirrus pixels. We divide the source-masked pixels by intensity intervals. For each interval, we compare the noise-corrected observed variance with the variance of the synthetic structure. The amplitude factors are interpolated according to the foreground brightness to produce a spatially varying amplitude map. The resulting texture follows the observed brightness-dependent fluctuation strength and preferred orientations, while retaining independent random phases.

\begin{figure}[!htbp]
\centering
  \resizebox{\hsize}{!}{\includegraphics{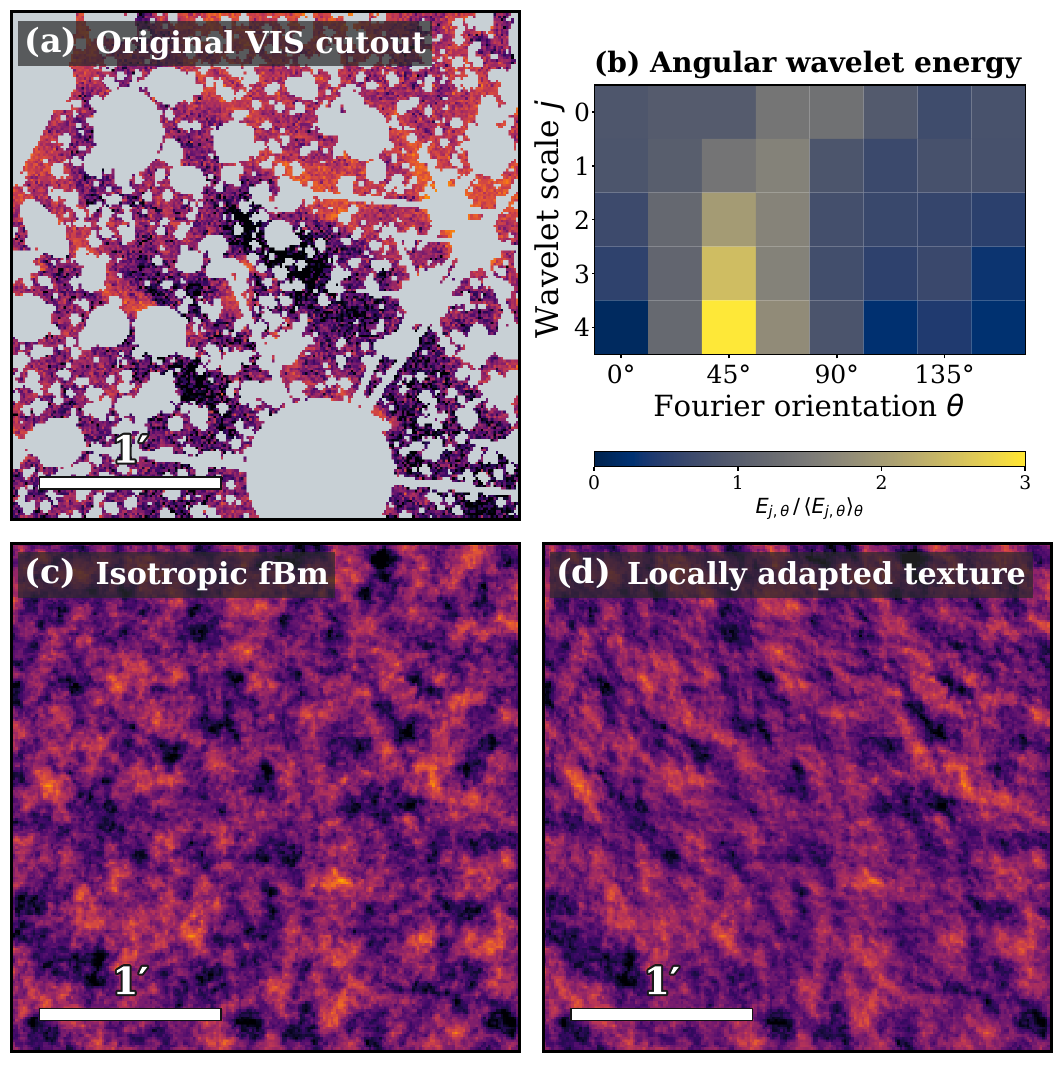}}
    \caption{
    Construction of locally constrained stochastic cirrus textures.
    (a) Source-masked image cutout example.
    (b) Angular wavelet energy measured across wavelet spatial scales and orientations, normalized by the mean energy at each scale. The directional contrast traces the preferred local orientation and strength of anisotropy.
    (c) Isotropic fBm texture generated with the assumed power-law spectrum.
    (d) The same realization after introducing scale-dependent anisotropy and brightness-dependent fluctuation measured from the observed pixels.
    The two synthetic textures share the same random phases.}
\label{fig:app_texture_modulation}
\end{figure}

Figure~\ref{fig:app_texture_modulation} illustrates how the stochastic texture is modulated. The measured angular wavelet energies provide scale-dependent directional information, which is used to transform an initially isotropic fBm realization into a locally anisotropic field. The two synthetic maps share the same random phases, while the modulated texture follows the preferred orientations and fluctuation strengths measured from the observed cirrus pixels.

The local stochastic texture $T^m$ is combined with the conditional broad field through the Gaussian scale separation in
Eq.~(\ref{eq:cirrus_initial}).
The resulting maps serve as starting points for the subsequent
optimization described in Sec.~\ref{sec:method_objective}.

\begin{figure}[!htbp]
\centering
  \resizebox{\hsize}{!}{\includegraphics{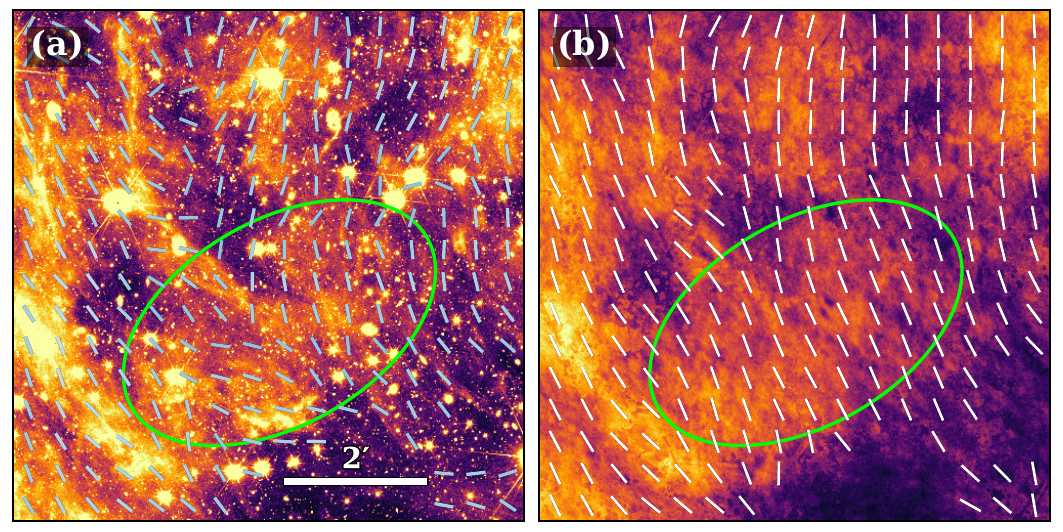}}
  \caption{Local filament orientations in the halo injection–recovery experiment. (a) Original \textit{Euclid} VIS image of Cutout-1 before halo injection. Blue sticks show orientations measured from the source-masked image. (b) Recovered cirrus from one realization. White sticks show  orientations measured from the recovered cirrus. Both use wavelet scale $j=0$, with local angular-energy pooling over approximately $1\arcmin$ FWHM. The green ellipse indicates the injection mask within which orientation information is excluded.}
\label{fig:app_orientation}
\end{figure}

Figure~\ref{fig:app_orientation} compares the local filament
orientations measured from the source-masked VIS image before halo
injection with those of one reconstructed cirrus realization after halo injection.
The reconstruction broadly follows the observed directional patterns where cirrus is bright, although individual filaments may not exactly match. Inside the target region, the local filament orientation is less constrained.

\section{Statistical constraints and numerical reconstruction}
\label{app:method}

\subsection{Multiscale statistical constraints}
\label{app:method_stats}

The WST constraints use five spatial scales and eight orientations. We combine the normalized first- and second-order scattering coefficient losses as the average of $S_1$ and $S_2$ losses:
\begin{equation}
\mathcal{L}_{\rm WST}
=
\frac{1}{2}
\left(
\mathcal{L}_{S_1}+\mathcal{L}_{S_2}
\right).
\end{equation}

The observed statistics are empirically corrected for the effects of source masking and correlated instrumental noise.
The mask response is calibrated using independent synthetic reference fBm fields, while noise contributions are estimated from eight instrumental-noise realizations.
These corrections are applied to the observed coefficients before constructing the statistical targets. 
When the targets are refreshed, the observed statistics are measured on the original mask support; the reconstructed pixels do not enter the measurements.

At the smallest wavelet scales, some coefficients are poorly constrained by the observations. We combine their corrected measurements with the corresponding statistics of the independent synthetic reference fBm fields:
\begin{equation}
S_i^{\rm target} =
\alpha_i S_i^{\rm obs} +
(1-\alpha_i)S_i^{\rm ref},
\label{eq:app_hybrid_wst}
\end{equation}
where the weight $\alpha_i$ is determined by the squared S/N of the observed coefficient $S_i$. The reference contribution becomes stronger where the observed cirrus signal is weak relative to its uncertainty.
This blending is applied to a selection of small-scale coefficients, including $S_1$ at $j=0,1$ and $S_2$ at $(j_1,j_2)=(0,1)$. Other scattering coefficients
retain their observation-based targets. The blended target uncertainties also include the dispersion of the reference realizations.

The $P_{00}$ terms constrain the power-like wavelet energy: 
\begin{equation}
P_{00}(j,\theta) =
\left\langle
\left|I_{\rm cir}\star\psi_{j,\theta}\right|^2
\right\rangle.
\label{eq:app_p00}
\end{equation}
$P_{00}$ describes the fluctuation power at each spatial scale and orientation. Note that this term is different from $S_1$ (Eq.~\ref{eq:wst_s1}), which averages the wavelet amplitude.

For each coefficient set $s\in\{S_1,S_2,P_{00}\}$,
the objective is:
\begin{equation}
\mathcal{L}_{s} =
\frac{1}{N_s}
\sum_{i\in\mathcal{I}_s}
\frac{\left(s_i^{\rm cir}-s_i^{\rm target}\right)^2}{\mathbf{s}_{i}}.
\label{eq:app_stat_loss}
\end{equation}
where $i$ stands for the wavelet scale and
orientation indices, $\mathcal{I}_s$ contains the reliable coefficients and $N_s$ is the sample size.
The normalization $\mathbf{s}_{i}$ combines the mean squared
target amplitude and its uncertainty, averaged over
orientations at the same scale or scale pair.
We only retain coefficients whose target amplitudes exceed their uncertainties in the calculation.

The CDF constraint compares quantiles of the high-frequency intensity fluctuations, obtained by subtracting a Gaussian-smoothed cirrus realization. This constraint preserves the distribution of positive and negative fluctuations without requiring individual structures to match pixel by pixel.

The orientation constraint compares the normalized angular wavelet energies across eight directions (Appendix~\ref{app:local_texture}), weighted by the uncertainty of the observed anisotropy measurements. This constraint favors similar orientations and directional contrasts in the reconstructed and observed cirrus, without requiring individual filaments to occupy the same positions.

\subsection{FIR and broad-field likelihoods}
\label{app:method_FIR_lld}

The two FIR-informed constraints are evaluated differently: the backbone term compares beam-matched intensities directly, while the broad correspondence is evaluated in Fourier space.

For $\mathcal{L}_{\rm bb}$, we smooth the reconstructed optical cirrus to the effective FIR beam and compare the predicted intensities with the Herschel measurements converted to optical units using the local optical--FIR calibration. The likelihoods are calculated using cirrus regions with reliable optical mask support, after accounting for measurement and calibration uncertainties.
For $\mathcal{L}_{\rm broad}$, we evaluate the Fourier power of the optical--FIR discrepancy $\Delta_{\rm b}$ and weight it by the inverse Matérn power spectrum defined in Appendix~\ref{app:init_broad}.
We apply an additional Gaussian weighting in Fourier
space, with a scale parameter of $\sigma_{\rm b}=30$ pixels, to reduce the influence of this regularization at high spatial frequencies. This allows the fine cirrus structures to be constrained primarily by the optical morphology, while retaining the broad-field covariance constraint on larger scales.

\subsection{Optical pixel likelihood}
\label{app:method_optical_lld}

The optical likelihood constrains the cirrus model using observed pixels compared with the sum of cirrus model, diffuse light model, and sky estimates. Let $O'$ denote the source-masked pixel domain excluding the central core. 
The optical residual is:
\begin{equation}
\boldsymbol{\delta}_{\rm opt} =
\left[
I_{\rm obs} - I_{\rm cir} - \widehat{I}_{\rm diff} - \widehat{B}
\right]^{O'}.
\label{eq:app_optical_residual}
\end{equation}
The corresponding objective is given by:
\begin{equation}
\mathcal{L}_{\rm opt} =
\frac{
\boldsymbol{\delta}_{\rm opt}^{\mathsf T}
\boldsymbol{\Sigma}_{\rm opt}^{-1}
\boldsymbol{\delta}_{\rm opt}}
{N_{\rm ext}},
\label{eq:optical_likelihood}
\end{equation}
where $N_{\rm ext}$ is the number of source-masked exterior pixels in $O$ used to normalize the optical constraint.

The covariance $\boldsymbol{\Sigma}_{\rm opt}$ combines correlated instrumental noise with an additional diagonal variance allowance within the target region. This allowance accounts for residual differences between the cirrus and diffuse-light models that are not explained by instrumental noise alone.
It is estimated from the exterior residuals and held fixed during optimization. Pixels near the excluded core are further down-weighted.

\subsection{Optimization and isophote fitting}
\label{app:method_optimization}

Only the cirrus field is optimized by an \texttt{Adam} gradient-based optimizer. For Cutout-1 and the central Abell~2390 field,
we perform 150 optimization steps in three stages of 50 steps,
with learning rates of 0.005, 0.003, and 0.001, respectively. For Cutout-2, we perform 100 steps with a fixed learning rate
of 0.005. 
Every 10 steps, after cirrus correction the residual sky is re-estimated using a $3\sigma$-clipped mean in the adopted outer annulus, followed by an updated diffuse light isophote fitting.

The isophote fitting assumes a fixed source center. Ellipticity and orientation are fitted where sufficient pixels are available, where we assume smooth radial variations to avoid abrupt changes in the model geometry. Beyond the last reliable isophote, the geometry is held fixed.

The same framework can, in principle, model multiple diffuse halos in a field by fitting their light distributions simultaneously, at the expense of increased computational cost. In this work, we restrict the analysis to a single target of interest within each cutout.

\section{Further examples for exponential halo injections}
\label{app:injection}

\begin{figure*}[!htbp]
\centering
  \resizebox{0.94\hsize}{!}{\includegraphics{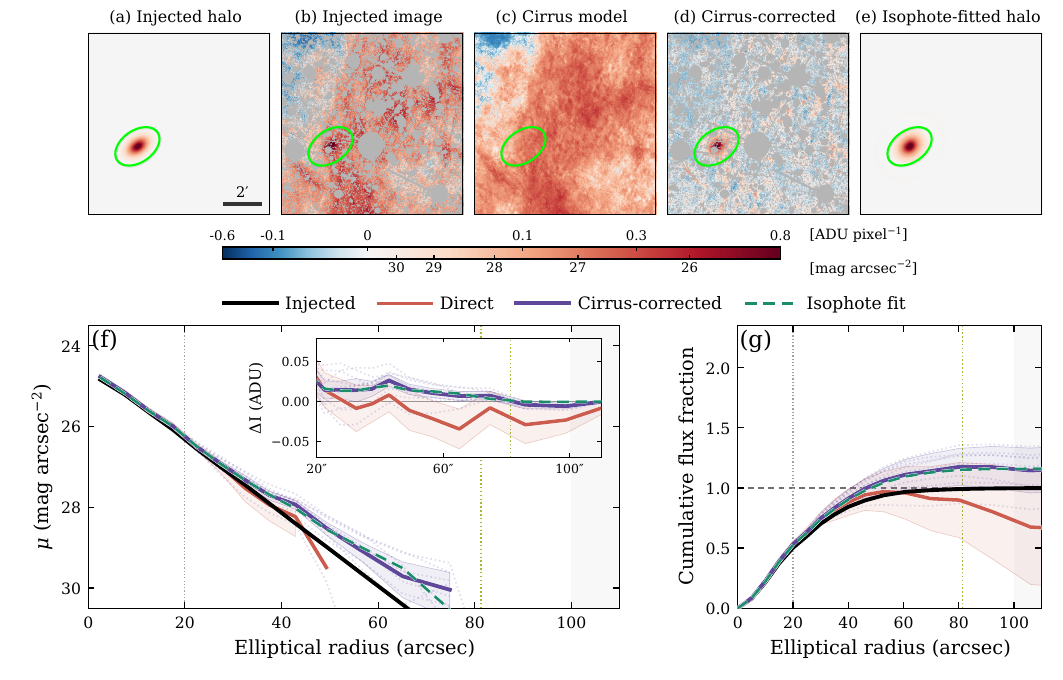}}
  \caption{Recovery of the injected exponential halo in Cutout-2. The layout is the same as in Fig.~\ref{fig:cutout2_injection_1}, with the same halo injected at a different position. Cirrus bias and sky bias cancel out in the direct profile. The corrected cumulative flux is marginally biased high.}
\label{fig:cutout2_injection2}
\end{figure*}

\begin{figure*}[!htbp]
\centering
  \resizebox{0.94\hsize}{!}{\includegraphics{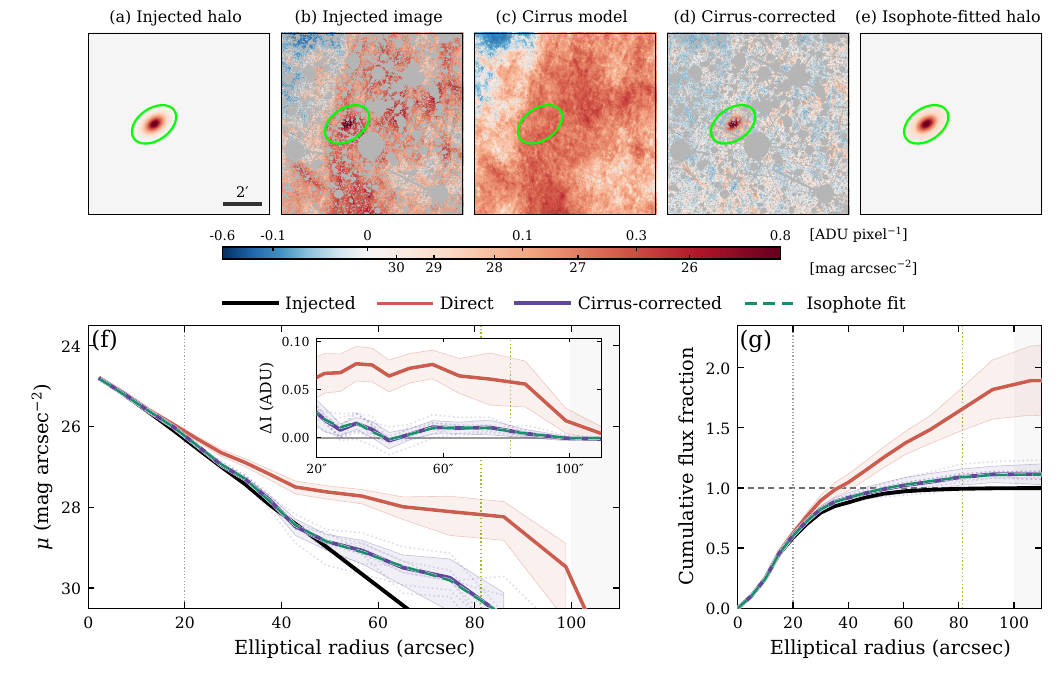}}
  \caption{Recovery of the injected exponential halo in Cutout-2. The layout is the same as in Fig.~\ref{fig:cutout2_injection_1}, with the same halo injected at a different position. This injection has a higher residual bias in the outskirts, while the bias in cumulative flux is much mitigated.}
\label{fig:cutout2_injection3}
\end{figure*}

Figures~\ref{fig:cutout2_injection2} and \ref{fig:cutout2_injection3} show two additional exponential halo injections in Cutout-2, both embedded within bright, structured cirrus. 
For Fig.~\ref{fig:cutout2_injection2}, the cirrus contamination and local sky overestimate partially cancel in the direct profile, producing apparently good agreement with the truth in 1D profile, despite that a strong 2D gradient persists across the target.
This case demonstrates that the measurement of diffuse emission is sensitive to the foreground and sky estimate. 
The cirrus-corrected profile remains slightly too bright beyond $\simeq 2R_{\rm e}$ ($40\arcsec$), with the cumulative flux overestimated by approximately 19\%. Larger deviation in cumulative flux through direct measurement, however, is observed at larger radii due to the overestimated sky. 

For Fig.~\ref{fig:cutout2_injection3}, the corrected profile presents a positive residual beyond $\simeq 2.5R_{\rm e}$. However, cirrus correction substantially reduces the light excess and the cumulative flux bias is reduced from approximately 67\% in the direct measurement to 10\% at $4R_{\rm e}$. These examples illustrate how the recovery depends on the local configuration of the target, cirrus, and sky, and how improvements in integrated photometry can coexist with residual biases in the light profile of the diffuse halo emission.

\section{Code and data availability}
\label{app:code}

The cirrus statistical modeling and reconstruction method developed in this work for diffuse halo light measurement is implemented in the Python package
\texttt{CirrusMachine}: \url{https://github.com/qliu4676/CirrusMachine}.
The repository also provides examples and data set for reproducing the halo light recovery experiment with the double de Vaucouleurs profile injection in Cutout-1 of the \textit{Euclid} ERO VIS Abell~2390 field.
      
\end{document}